\documentclass[iop]{emulateapj}
\pdfoutput=1

\usepackage{natbib}
\usepackage{amsmath,amsthm,amssymb}
\usepackage{setspace,multirow,hhline}
\usepackage{etex,hyphenat}
\usepackage{graphicx}
\DeclareGraphicsExtensions{.pdf}
\graphicspath{{}{graphics/}}

\usepackage{xcolor}
\usepackage{hyperref}
\hypersetup{
   colorlinks,
   linkcolor={blue!88!black!80},
   citecolor={blue!88!black!80},
   urlcolor={blue!88!black!80}
}

\usepackage{bm}
\usepackage[T1]{fontenc}      
\newcommand{\ditto}[1][.4pt]{\textquotedbl}

\newcommand{\Lya}{Ly$\alpha$}
\newcommand{\OII}{[O\,\textsc{ii}]}

\newcommand{\NeIII}{[Ne\,\textsc{iii}]}
\newcommand{\Hb}{H$\beta$}
\newcommand{\OIII}{[O\,\textsc{iii}]}
\newcommand{\CIII}{C\,\textsc{iii}]}
\newcommand{\CIV}{C\,\textsc{iv}}
\newcommand{\MgII}{Mg\,\textsc{ii}}
\newcommand{\Ha}{H$\alpha$}

\newcommand{\Paa}{Pa$\alpha$}
\newcommand{\Pab}{Pa$\beta$}
\newcommand{\FeII}{[Fe\,\textsc{ii}]}
\newcommand{\EWLya}{$W_{\rm{Ly}\alpha}$}
\newcommand{\EWobs}{$\rm{EW}_{\rm{obs}}$}
\newcommand{\wlobs}{$\lambda_{_{\rm EL}}$}
\newcommand{\dA}{$d_A$}
\newcommand{\HETDEX}{\nohyphens{HETDEX}}

\shorttitle{Bayesian Classification of Emission-line Galaxies}
\shortauthors{Leung, Acquaviva, Gawiser, et al.}
\slugcomment{Submitted to ApJ}

\begin{document}

\title{Bayesian Redshift Classification of Emission-line Galaxies with Photometric Equivalent Widths}
\author{
  Andrew S. Leung\altaffilmark{1,2}, 
  Viviana Acquaviva\altaffilmark{3},
  Eric Gawiser\altaffilmark{1},
  Robin Ciardullo\altaffilmark{4,5},
  Eiichiro Komatsu\altaffilmark{6,7},\\
  A.I. Malz\altaffilmark{8},
  Gregory R. Zeimann\altaffilmark{4,5,2},
  Joanna S. Bridge\altaffilmark{4,5},
  Niv Drory\altaffilmark{9},
  John J. Feldmeier\altaffilmark{10},\\
  Steven L. Finkelstein\altaffilmark{2},
  Karl Gebhardt\altaffilmark{2},
  Caryl Gronwall\altaffilmark{4,5},
  Alex Hagen\altaffilmark{4,5},\\
  Gary J. Hill\altaffilmark{2,9}, and
  Donald P. Schneider\altaffilmark{4,5}
}
\altaffiltext{1}{Department of Physics and Astronomy,
  Rutgers, The State University of New Jersey, Piscataway, New Jersey 08854, USA; 
  \href{mailto:gawiser@physics.rutgers.edu}{\tt gawiser@physics.rutgers.edu}}
\altaffiltext{2}{Department of Astronomy, 
  The University of Texas at Austin, Austin, Texas 78712, USA;
  \href{mailto:leung@astro.as.utexas.edu}{\tt leung@astro.as.utexas.edu}}
\altaffiltext{3}{Department of Physics, New York City College of Technology,
  The City University of New York, Brooklyn, New York 11201, USA; 
  \href{mailto:vacquaviva@citytech.cuny.edu}{\tt vacquaviva@citytech.cuny.edu}}
\altaffiltext{4}{Department of Astronomy and Astrophysics, 
  The Pennsylvania State University, University Park, Pennsylvania 16802, USA}
\altaffiltext{5}{Institute for Gravitation and the Cosmos, 
  The Pennsylvania State University, University Park, Pennsylvania 16802, USA}
\altaffiltext{6}{Max-Planck-Institut f\"{u}r Astrophysik,
  85740 Garching bei M\"{u}nchen, Germany}
\altaffiltext{7}{Kavli Institute for the Physics and Mathematics of the Universe (Kavli IPMU, WPI), 
  Todai Institutes for Advanced Study, The University of Tokyo, 
  Kashiwa 277-8583, Japan}
\altaffiltext{8}{Center for Cosmology and Particle Physics, Department of Physics, 
  New York University, New York, New York 10003, USA}
\altaffiltext{9}{McDonald Observatory, 
  The University of Texas at Austin, Austin, Texas 78712, USA}
\altaffiltext{10}{Department of Physics and Astronomy, 
  Youngstown State University, Youngstown, Ohio 44555, USA}

\begin{abstract}
We present a Bayesian approach to the redshift classification of emission-line galaxies when 
only a single emission line is detected spectroscopically. We consider the case of surveys 
for high-redshift \Lya-emitting galaxies (LAEs), which have traditionally been classified 
via an inferred rest-frame equivalent width (\EWLya) greater than 20\,\AA. Our Bayesian method 
relies on known prior probabilities in measured emission-line luminosity functions and equivalent 
width distributions for the galaxy populations, and returns the probability that an object in 
question is an LAE given the characteristics observed. This approach will be directly relevant 
for the Hobby--Eberly Telescope Dark Energy Experiment (\HETDEX), which seeks to classify 
$\sim$\,$10^6$ emission-line galaxies into LAEs and low-redshift \OII~emitters. For a simulated 
\HETDEX\ catalog with realistic measurement noise, our Bayesian method recovers 86\,\% of LAEs 
missed by the traditional \EWLya\,$>$\,20\,\AA\ cutoff over 2\,$<$\,$z$\,$<$\,3, outperforming 
the equivalent width (EW) cut in both contamination and incompleteness. This is due to the method's 
ability to trade off between the two types of binary classification error by adjusting the stringency 
of the probability requirement for classifying an observed object as an LAE. In our simulations of 
\HETDEX, this method reduces the uncertainty in cosmological distance measurements by 14\,\% with 
respect to the EW cut, equivalent to recovering 29\,\% more cosmological information. Rather than 
using binary object labels, this method enables the use of classification probabilities in 
large-scale structure analyses. It can be applied to narrowband emission-line surveys as well as 
upcoming large spectroscopic surveys including {\it Euclid} and WFIRST.
\vspace{2pt}
\end{abstract}


\section{Introduction}
\label{sec:intro}
The Hobby--Eberly Telescope Dark Energy Experiment\footnote{\url{http://www.hetdex.org}} (\HETDEX) 
will obtain redshifts for approximately a million \Lya-emitting galaxies (LAEs) in its upcoming 
wide-field survey and to determine the local Hubble expansion rate, angular diameter distance (\dA), 
and the growth of structure 
in the 1.9\,$<$\,$z$\,$<$\,3.5 universe \citep{Hill+08}. To achieve its science goals, \HETDEX\ 
requires an accurate classifier to identify LAEs and discard contaminants (primarily $z$\,$<$\,0.5 
\OII~emitters) from the statistical LAE sample.

\Lya\ is a spectral feature produced by the transition of neutral hydrogen atoms from the first 
excited state to the ground state ($n$\,=\,2\,$\rightarrow$\,1), with a rest-frame wavelength of 
1216\,\AA. The Visible Integral-field Replicable Unit Spectrograph (VIRUS), currently being 
deployed at the Hobby--Eberly Telescope (HET), has a spectral range of 3500--5500\,\AA\ 
\citep{Hill+15}, making spectroscopic detections of \Lya\ possible between 
redshifts 1.88\,$<$\,$z$\,$<$\,3.52. This capability enables the primary science goal of the 
\HETDEX\ survey, the measurement of the expansion history of the universe \citep{Hill+08, Adams+11}, 
by constraining cosmological parameters from the observed signature of baryon acoustic oscillations 
(BAO) in the power spectrum of LAE redshifts and positions \citep{Seo+Eisenstein03, 
Blake+Glazebrook03, Hu+Haiman03, Seo+Eisenstein07, Koehler+07, Shoji+09}. \Lya\ is usually the
strongest line, therefore near the survey limit, where most of the targets lie, the VIRUS spectra 
will have only one detected emission line.

Another prominent emission line feature in galaxy spectra is the \OII\ doublet at $\sim$\,3726\,\AA\ 
and $\sim$\,3729\,\AA, a pair of atomic transitions for the decay of singly ionized oxygen. Galaxies 
with strong \OII\ emission from $z$\,$<$\,0.476 will also be identified by \HETDEX\ as a single-line 
detection within its spectral range, as the \OII\ $\lambda$\,3727 doublet is separated by 2.5\,\AA\ 
\citep[rest frame;][]{Osterbrock74} and cannot be resolved by the VIRUS instrument (spectral 
resolution: 5.7\,\AA). Distinguishing LAEs targeted by \HETDEX\ from low-redshift \OII~emitters 
therefore represents a challenge in a blind spectroscopic survey when only one emission line is 
detected \citep{Adams+11}.

\begin{figure}
\centering
\includegraphics[width=\linewidth,trim={0.1in 0.06in 0.12in 0.0in},clip]
{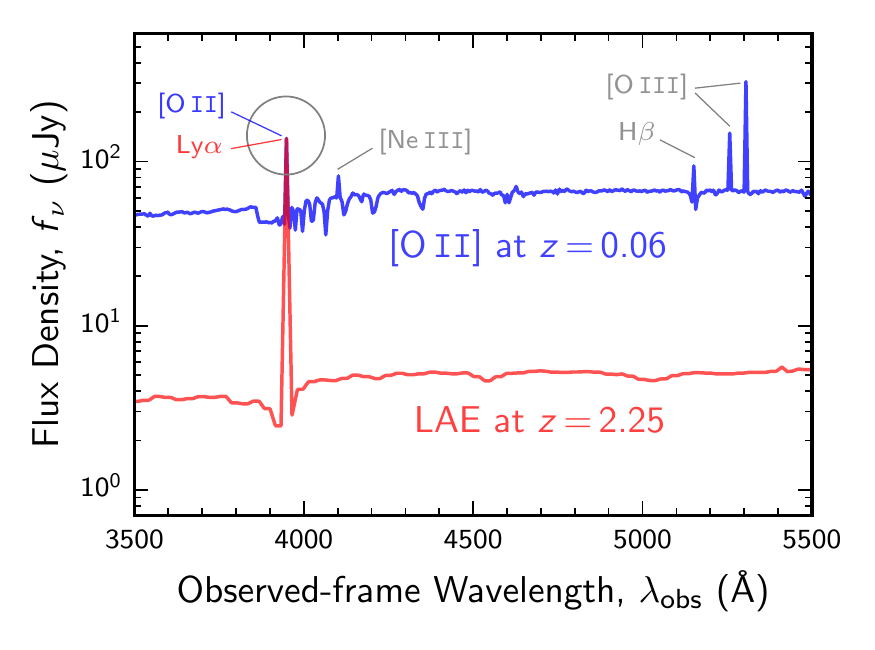}
\caption{
The spectra of two model galaxies over the spectral range of \HETDEX\ are shown 
in this example. The \Lya\ line in the spectrum of the high-redshift galaxy 
(red) is observed at the same wavelength as the unresolved \OII\ 
doublet from a low-redshift galaxy (blue) in the survey foreground. Intensity 
of the primary emission line (indicated by the gray circle) is identical for 
the two galaxies depicted, but the continuum flux densities of the two 
spectra are different. The LAE in this example, as is typically the case, has 
a much larger equivalent width than the \OII\ emitter depicted. The use of \NeIII, 
\Hb\ and \OIII\ lines in applicable cases as additional spectral information for 
Bayesian classification is discussed in \S\,\ref{subsec:results-aper-spec}.\\
}
\label{fig:spectra-example}
\end{figure}

Figure \ref{fig:spectra-example} illustrates this classification problem. In the example, a 
high-redshift LAE and a foreground \OII\ emitter are detected via their primary emission lines 
observed at the same wavelength, with identical fluxes measured for the detected lines. In order 
to make an accurate redshift classification, we need to consider additional information available 
in our measurements. narrowband selection for strong \Lya\ emission in the literature typically 
requires LAEs to have rest-frame equivalent width (\EWLya) greater than 20\,\AA\ 
\citep[e.g.,][]{Cowie+Hu98, Gronwall+07}. This equivalent width (EW) method is effective in 
limiting the misclassification of \OII~emitters as LAEs \citep{Gawiser+06} at 
redshift 2\,$<$\,$z$\,$<$\,3, at the expense of having an incomplete sample of LAEs. However, 
fractional contamination in the LAE sample increases rapidly with redshift due to two concurrent 
factors: the rapid rise in the volume occupied by \OII~emitters and the increase in their average 
intrinsic equivalent widths \citep{Hogg+98, Ciardullo+13}. Using the simple EW cut reduces the 
purity and completeness of a statistical sample of objects classified as LAEs and thus the 
precision with which \HETDEX\ will be able to measure the evolution of dark energy \citep{Komatsu10}.

The distributions of line luminosity and equivalent widths have been measured for emission-line 
selected samples of LAEs \citep[e.g.,][]{Shimasaku+06, Gronwall+07, Ouchi+08, Guaita+10, Ciardullo+12} 
and \OII~emitters \citep{Ciardullo+13} over the spectral range of the \HETDEX\ survey. In addition, 
the equivalent width distribution of \OII~emitters in the local universe has been extremely well 
measured using both continuum and emission-line selected galaxy samples \citep{Blanton+Lin00, 
Gallego+96, Gallego+02}. With these distribution functions as prior probabilities, we can compute 
the relative likelihood that a detected emission-line object is an LAE or an \OII\ emitter given its 
observed characteristics.

We explore this Bayesian approach to classifying emission-line galaxies as a means to improve the 
quality of the cosmological sample of LAEs. Section \ref{sec:methodology} describes our methodology, 
including details of the simulation of a \HETDEX\ catalog consisting of LAEs and \OII~emitters 
(\S\,\ref{sec:simulate}) and an overview of the statistical framework for a Bayesian method that 
can be used to identify LAEs in a line flux-limited sample of emission-line galaxies 
(\S\,\ref{sec:bm}). Section \ref{sec:results} presents the results of Bayesian classification of 
LAEs in a simulated \HETDEX\ catalog and quantifies the improvement over the equivalent width method. 
Section \ref{sec:disc} offers a discussion of our findings and their applications.

Throughout the present work, we assume a $\Lambda$CDM cosmology with $H_0$~=~70~km\,s$^{-1}$\,Mpc$^{-1}$, 
$\Omega_m$~=~0.3, and $\Omega_{\Lambda}$~=~0.7 \citep{Komatsu+11,Planck15}. All magnitudes are 
reported in the AB system \citep{Oke+Gunn83}.

\section{Methodology}
\label{sec:methodology}

\subsection{Simulated Catalog of Emission-line Galaxies}
\label{sec:simulate}
We simulate populations of LAEs and \OII~emitters on which to test the methods for galaxy 
classification. For the simulations, we specify a 300~deg$^2$
survey area (the size of the \HETDEX\ spring field, roughly two-thirds of the total survey area) 
and a 1/4.5 filling factor to mimic the design of the upcoming \HETDEX\ survey \citep{Hill+15}.

\subsubsection{Spectroscopic Survey Simulation}
\label{subsec:sim-spec}
\begin{figure*}
\centering
\includegraphics[width=\linewidth,trim={0.2in 0.12in 0.2in 0.12in},clip]
{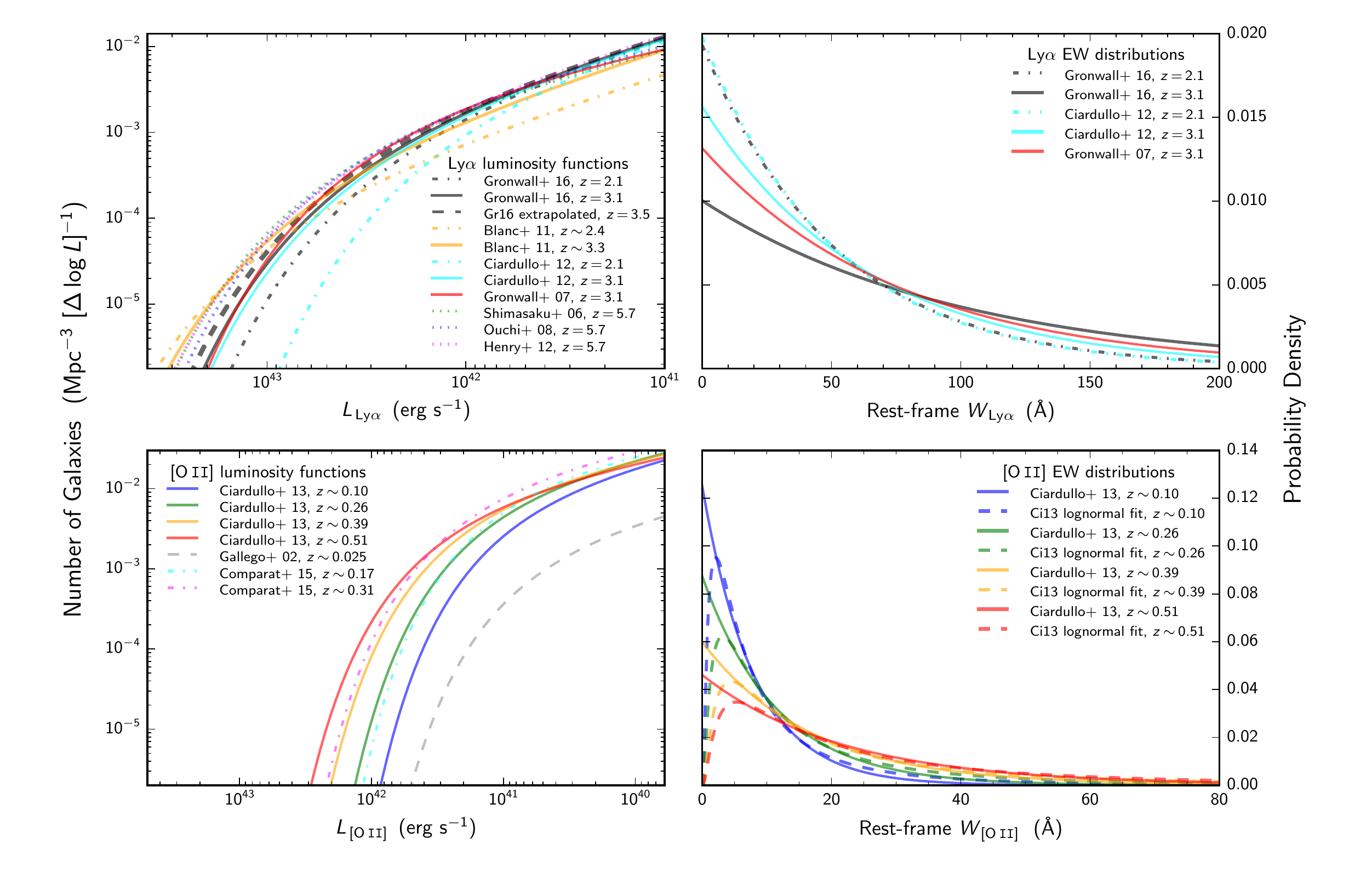}
\caption{
Luminosity functions and equivalent width distributions for LAEs and 
\OII\ emitters. 
\textit{Upper left}: The dashed black curve is obtained by linear 
extrapolation of Schechter function parameters reported by \citet{Gronwall+16}
out to $z$\,=\,3.5; it is consistent with \Lya\ luminosity functions measured 
at higher redshifts \citep{Shimasaku+06, Ouchi+08, Henry+12}. 
\textit{Upper right}: \Lya\ EW distributions assumed in simulations in the 
present work; \citet{Gronwall+16} parameters evolving linearly with redshift 
constitute our ``baseline'' scenario (see Table~\ref{tab:compare}). 
\textit{Lower right}: Dashed lines show lognormal fits at EW\,$>$\,5\,\AA\ to 
the exponential equivalent width distributions measured by \citet{Ciardullo+13} 
(see \S\,\ref{subsec:disc-imperfect}); the exponential form is used in simulations 
of \OII\ equivalent widths in the present work. 
\textit{Lower left}: \OII\ luminosity functions used in our simulations 
\citep{Ciardullo+13} are consistent with published results at similar redshifts 
\citep{Gallego+02, Comparat+15}.\\
}
\label{fig:sim_param}
\end{figure*}

\citet{Gronwall+07}, \citet{Ciardullo+12}, and \citet[][in prep; hereafter \texttt{Gr16}]{Gronwall+16} 
have measured the luminosity functions for LAE populations at $z$\,=\,2.1 and $z$\,=\,3.1. Using 
the {\tt Gr16} luminosity functions, we simulate \Lya\ line luminosities via Monte Carlo simulations. 
We use a \citet{Schechter76} function of the form:
\begin{equation}
  \Phi(L)\,dL = \phi^{*}\,(L/L^{*})^{\alpha}\,\exp(-L/L^{*})\,d(L/L^{*}) \,,
\label{eq:schechter}
\end{equation}
and assume that the parameters (shown in Table~\ref{tab:sim_param}) evolve linearly with redshift. 
We obtain the distribution parameters at redshifts 2.1\,$<$\,$z$\,$<$\,3.1 by linear interpolation 
of the \texttt{Gr16} parameter values for $z$\,=\,2.1 and $z$\,=\,3.1; for simulated LAEs at 
$z$\,$<$\,2.1 and $z$\,$>$\,3.1, we linearly extrapolate the parameters. The top left panel in 
Figure~\ref{fig:sim_param} shows that our extrapolation of the \texttt{Gr16} luminosity function 
to $z$\,=\,3.5 is consistent with the weakly evolving \Lya\ luminosity functions measured at higher 
redshift \citep{Shimasaku+06, Ouchi+08, Henry+12}.

The measured equivalent width distributions of LAEs have been modeled with various distributions. 
For our simulations, we assume an exponential form with a scale length, $w_{_0}$:
\begin{equation}
  \Psi(W)\,dW = \frac{e^{_{-W/w_{_0}}}}{w_{_0}}\,dW \,.
\label{eq:exponential}
\end{equation}
As with the Schechter function parameters, we assume the exponential scale lengths evolve linearly 
with redshift. Since previous studies have found weak or no correlation between emission-line 
luminosity and equivalent width \citep{Cowie+96, Hogg+98, Gronwall+07, Ciardullo+12, Ciardullo+13}, 
we model the luminosity function and the equivalent width distribution as orthogonal functions.

\begin{table*}
\caption{
Schechter function parameters and exponential scale lengths for LAEs 
and \OII~emitters;\\parameters for fitted lognormal \OII\ equivalent
width distributions are shown in the far-right column
\label{tab:sim_param}
}
\vspace{2pt}
\centering
\scriptsize
\begin{spacing}{1.45}
\begin{tabular}{ccccccccccc}
\hhline{===========}

    &&&& \multicolumn{3}{c}{Schechter function}
    &
    & {Exponential EW}
    &
    & \multicolumn{1}{c}{Lognormal EW$^{\rm \,a,b}$} \\

\hhline{~~~~---~-~-}
\multicolumn{3}{c}{Sample} 
    && $\alpha$
    & $\log L^{*}$
    & $\log \phi^{*}$
    &
    & $w_{_0}$
    &
    & $W_{_0}$ 
    \\

    &&&&
    & (erg\,s$^{-1}$)
    & (Mpc$^{-3}$)
    && (\AA)
    && (\AA)
    \\

\cline{1-11}

\ LAE\ \ & $z$\,$\sim$\,2.1 & \texttt{Gr16}$^{\rm \,c}$
    && $-1.65$
    & $42.61^{+0.17}_{-0.06}$
    & $-3.08^{+0.04}_{-0.06}$
    && $50^{+5}_{-4}$
    && --- \\
\ditto & \ditto & \texttt{Ci12}$^{\rm \,d}$
    && $-1.65$
    & $42.33^{+0.12}_{-0.12}$
    & $-2.86^{+0.05}_{-0.05}$
    && $50^{+7}_{-7}$
    && --- \\

\cline{1-11}

LAE & $z$\,$\sim$\,3.1 & \texttt{Gr16}\ \
    && $-1.65$
    & $42.77^{+0.10}_{-0.08}$
    & $-2.98^{+0.07}_{-0.06}$
    && $100^{+28}_{-14}$
    && --- \\
\ditto & \ditto & \texttt{Ci12}\ \
    && $-1.65$
    & $42.76^{+0.10}_{-0.10}$
    & $-3.17^{+0.05}_{-0.05}$
    && $64^{+9}_{-9}$
    && --- \\
\ditto & \ditto & \texttt{Gr07}$^{\rm \,e}$
    && $-1.36$
    & $42.66^{+0.26}_{-0.15}$
    & $-2.89^{+0.04}_{-0.03}$
    && $76^{+11}_{-8}$
    && --- \\

\cline{1-11}

\OII & $z$\,$\sim$\,0.10 & \texttt{Ci13}$^{\rm \,f}$
    && $-1.20$
    & $41.07^{+0.18}_{-0.16}$
    & $-2.34^{+0.05}_{-0.06}$
    && $8.0^{+1.6}_{-1.6}$
    && $3.8^{+0.9}_{-0.8}$
    \\

\ditto & $z$\,$\sim$\,0.26 & \ditto
    && $-1.20$
    & $41.29^{+0.11}_{-0.11}$
    & $-2.34^{+0.03}_{-0.03}$
    && $11.5^{+1.6}_{-1.6}$
    && $5.8^{+0.8}_{-0.9}$
    \\

\ditto & $z$\,$\sim$\,0.39 & \ditto
    && $-1.20$
    & $41.50^{+0.08}_{-0.10}$
    & $-2.43^{+0.01}_{-0.01}$
    && $16.6^{+2.4}_{-2.4}$
    && $8.2^{+0.9}_{-1.0}$
    \\

\ditto & $z$\,$\sim$\,0.51 & \ditto
    && $-1.20$
    & $41.68^{+0.08}_{-0.12}$
    & $-2.54^{+0.01}_{-0.01}$
    && $21.5^{+3.3}_{-3.3}$
    && $10.1^{+1.1}_{-1.2}$
    \\

\cline{1-11}
\multicolumn{11}{l}{$^{\rm a}$\,\hspace{0.55pt}Lognormal distribution parameter
   $\sigma_{_W}$ is fixed at 1.1 for all redshifts $z_{\rm \,\OII}$;
   see Equation~(\ref{eq:lognormal}).} \vspace{-3pt}\\
\multicolumn{9}{l}{$^{\rm b}$\,\hspace{0.05pt}The consequences of using lognormally distributed \OII\ EWs are discussed in \S\,\ref{subsec:disc-imperfect}.}&&\vspace{-3pt}\\
\multicolumn{3}{l}{$^{\rm c}$\,\hspace{0.6pt}\citet{Gronwall+16}} &&&&&&&& \vspace{-3pt}\\
\multicolumn{3}{l}{$^{\rm d}$\,\citet{Ciardullo+12}} &&&&&&&& \vspace{-3pt}\\
\multicolumn{3}{l}{$^{\rm e}$\,\hspace{0.6pt}\citet{Gronwall+07}} &&&&&&&& \vspace{-3pt}\\
\multicolumn{3}{l}{$^{\rm f}$\,\hspace{0.88pt}\citet{Ciardullo+13}} &&&&&&&& \vspace{3pt}\\
\end{tabular}
\end{spacing}
\end{table*}

The simulation of \OII\ luminosities follows the same procedure as described above, with Schechter 
function parameters reported by \citet[][hereafter \texttt{Ci13}]{Ciardullo+13}. The \OII\ luminosity 
function has also been measured in studies by \citet{Gallego+02}, \citet{Teplitz+03}, 
\citet{Hippelein+03}, \citet{Ly+07}, \citet{Takahashi+07}, and \citet{Comparat+15}. For simplicity, 
we assume the exponential form in Equation~(\ref{eq:exponential}) for the \OII\ equivalent width 
distributions, with {\tt Ci13} parameters linearly evolving with redshift. 

To model the evolution of the galaxy property distributions, we choose 
small values of $\Delta{z}$ for our simulations ($\Delta{z}$\,=\,0.01 for 
LAEs; and $\Delta{z}$\,=\,0.005 for \OII~emitters, whose corresponding 
volume elements are smaller). The total number of each type of object to 
be simulated in a given redshift bin is given by the product of the
comoving volume of the survey area in $\Delta{z}$ and the integral of 
the Schechter function above a conservative minimum flux, below which
there can be no line detections on VIRUS. For each simulated object, 
the realized redshift and object type (LAE or \OII\ emitter) determine 
the observed-frame wavelength ($\lambda_{_{\rm{EL}}}$) of the primary 
simulated emission line. Since \Lya\ luminosity functions were measured 
for LAEs selected with the \EWLya\,$>$\,20\,\AA\ requirement 
\citep[e.g.,][]{Ciardullo+12, Gronwall+16} and our Monte Carlo
simulations realize equivalent widths from probability distributions 
that go down to 0\,\AA, we must correct the number density of LAEs to 
account for the fraction that would have \EWLya\,$\leq$\,20\,\AA.\footnote{The
   correction factor is equal to $1/e^{-20{\textup{\AA}}/w_{_0}}$. Assuming the 
   results in \texttt{Gr16}, the factors at $z$\,=\,2.1 and $z$\,=\,3.1 are 1.49 
   and 1.22, respectively.}

Other emission lines are added to the simulated spectra of \OII\
emitters, based on relative line strengths as a function of
metallicity \citep{Anders+Fritze03}, which we assume to be
one-fifth of solar for low-redshift objects in our simulations. 
The nearest \OII~emitters ($z$\,$\lesssim$\,0.1) have four 
other strong emission lines\,---\,\NeIII\,$\lambda$3869, 
\Hb\,$\lambda$4861, \OIII\,$\lambda$4959, and
\OIII\,$\lambda$5007\,---\,that may be detected within the spectral
range of \HETDEX\ (see Figure~\ref{fig:spectra-example}), depending on
the redshift.

Gaussian noise with a mean equal to zero and a standard deviation equal to
one-fifth of the wavelength-dependent $5\sigma$ line flux sensitivity 
limit of \HETDEX\ is added to simulated line fluxes. Subsequent to 
the addition of noise, simulated objects with ``recorded'' line fluxes 
that fall below the $5\sigma$ detection limit are eliminated from the 
``observable'' sample, resulting in a $5\sigma$ line flux-limited 
sample of emission-line galaxies for our simulated catalog.

\subsubsection{Imaging Survey Simulation}
\label{subsec:sim-phot}
We explore a scenario specific to our application (\HETDEX), where
spectroscopic line fluxes are coupled with continuum flux densities 
obtained through aperture photometry on broadband imaging, resulting 
in measurements of \textit{photometric} equivalent widths. Noiseless 
simulated emission line fluxes ($f_{_{\rm{EL}}}$) and observed-frame
equivalent widths (\EWobs) are converted into continuum flux densities
($f_{\nu\rm{,\,cont}}$) at the observed emission-line wavelength, 
as follows:
\begin{eqnarray}
\label{eq:EWdef}
  \rm{EW}_{\rm{obs}} \, 
  &=& \, \dfrac{f_{_{\rm{EL}}}}{f_{\lambda\rm{,\,cont}}} \,,\\
\label{eq:f_nudef}
  f_{\nu\rm{,\,cont}}(\lambda_{_{\rm{EL}}}) \,
  &=& \, f_{\lambda\rm{,\,cont}} \, \dfrac{\lambda_{_{\rm{EL}}}^{\,2}}{c} \,.
\end{eqnarray}

For each galaxy in the $5\sigma$ line flux-limited sample, we simulate an
imaging survey counterpart by extrapolating its continuum flux density
from the observed emission line wavelength ($\lambda_{_{\rm{EL}}}$)
to a specified broadband imaging survey filter. The power law slope of
the continuum is simulated from the distributions of $r'$$-$$z'$ colors
of LAEs and \OII~emitters observed in the \HETDEX\ Pilot Survey\footnote{A 
   Gaussian fit of the $r'$$-$$z'$ colors of LAEs in the HPS sample has mean 
   $\mu$\,=\,0.11 and standard deviation $\sigma$\,=\,0.55; for \OII~emitters 
   in HPS, a Gaussian fit of $r'$$-$$z'$ has $\mu$\,=\,0.53 and $\sigma$\,=\,0.28.} 
\citep[HPS;][]{Adams+11, Blanc+11, Bridge+15}. The procedure
prescribed by \citet{Madau95} is applied to simulated spectra as a 
function of observed wavelength to correct for absorption by neutral 
hydrogen in the intergalactic medium \citep[the ``\Lya\ forest'';][]{Lynds71}. 
The sum of the resulting flux from the continuum and flux contributed by emission 
lines, multiplied by the transmission fraction of the specified imaging filter 
(including the quantum efficiency of the CCD) results in the continuum flux 
density observed for each galaxy in simulated aperture photometry. One-fifth
of the $5\sigma$ depth of the simulated imaging survey \citep[SDSS $g'$ 
and $r'$ filters are assumed in this work;][]{Doi+10} is used as the standard 
deviation for the Gaussian profile of measurement noise in photometry.

For each model $5\sigma$ emission line detection, we compute a new 
quantity\,---\,{\it photometric equivalent width}\,---\,as the relative 
strength of the simulated line flux to the continuum flux density measured in 
aperture photometry. These simulated photometric equivalent widths are 
observed-frame quantities with measurement noise propagated from both 
simulated line flux and simulated continuum flux density measurements. 
As a result of their larger equivalent widths and greater luminosity 
distances from Earth, among $5\sigma$ line detections LAEs are generally 
fainter than \OII~emitters in their continua, and their measurements therefore
suffer from larger fractional uncertainties than continuum measurements 
of \OII~emitters.

\begin{figure}
\centering
\includegraphics[width=\linewidth,trim={0.02in 0.0in 0.06in 0.0in},clip]
{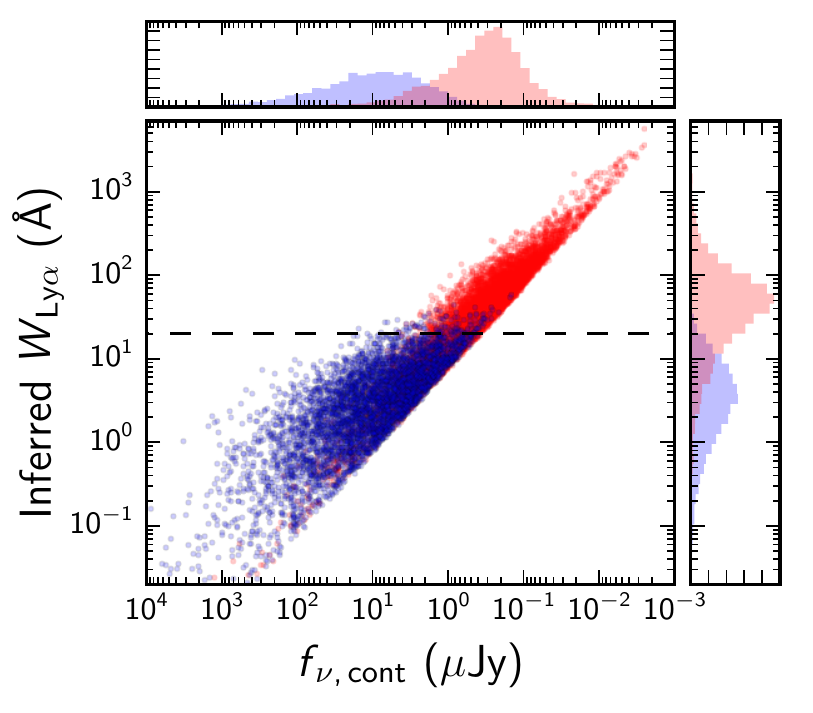}
\caption{
Simulated \HETDEX\ catalog of 1.4$\times$10$^6$ emission-line 
galaxies, with $g'$ band imaging ($5\sigma$\,=\,25.1) and realistic measurement
noise. Approximately 1\,\% of the simulated catalog is displayed in the central plot.
842,600 ``true'' LAEs (1.9\,$<$\,$z$\,$<$\,3.5) are shown in red;
538,300 \OII~emitters ($z$\,$<$\,0.48) are indicated in blue. The
horizontal dashed line marks
EW\,=\,20\,\AA\ in the emission rest frame of \Lya\ (the traditional 
selection requirement). The projected histogram along the vertical axis 
shows the distributions of equivalent widths of the simulated galaxies 
in the emission rest frame, inferred by assuming each primary line 
detection is \Lya. The projected histogram along the horizontal axis
presents the distributions of continuum density density measured for the 
galaxies in simulated aperture photometry.\\
}
\label{fig:sample-all}
\end{figure}

Figure \ref{fig:sample-all} presents the distribution of inferred
\Lya\ equivalent widths (\EWLya) for a $5\sigma$ line flux-limited
sample of simulated LAEs and corresponding \OII~emitters, plotted 
against their continuum flux densities ($f_{\nu\rm{,\,cont}}$) obtained in
simulated aperture photometry on $g'$ band imaging with $5\sigma$ 
depth of 25.1 mag. The horizontal dashed line denotes an 
equivalent width of 20\,\AA\ in the rest frame of \Lya\ emission.

\subsection{Bayesian Classification of Emission-line Galaxies}
\label{sec:bm}
\HETDEX\ will use the two-point correlation function, or power
spectrum, of high-redshift LAEs to measure dark energy evolution
over 1.9\,$<$\,$z$\,$<$\,3.5 \citep{Hill+08,Adams+11}. When \OII\ 
emitters are misidentified as LAEs, the correlations of low-redshift 
galaxies with smaller comoving separations are erroneously mapped to 
the correlation function of LAEs \citep{Komatsu10}. The observed 
power spectrum is given by the weighted average of the LAE power 
spectrum, $P_{\rm{\,LAE}}$, and the contamination power spectrum:\footnote{See 
Appendix~\ref{apndx:A} for the derivation of Equation (\ref{eq:pobscopy}).}
\begin{multline}
  P_{\rm{\,obs}}\left(k_\perp,k_\parallel\right)
  = {\left(1-f_{\rm \OII}\right)^2}\,
  P_{\rm{\,LAE}}\left(\sqrt{k_\perp^2+k_\parallel^2}\right) \\
  + {f_{\rm \OII}^2} ({\alpha^2\beta})\,
  P_{\rm{\,\OII}}\left(\sqrt{\alpha^2k_\perp^2+\beta^2k_\parallel^2}\right) \,,
\label{eq:pobscopy}
\end{multline}
where $f_{\rm \OII}$ is the fraction of \OII~emitters in the total
LAE sample, i.e., 
\begin{equation}
  f_{\rm \OII}\equiv \frac{\mbox{number of contaminating \OII\
  emitters}}{\mbox{number of galaxies classified as LAEs}} \,.
\label{eq:contamfrac}
\end{equation}

Improvement in the quality of the LAE sample can be achieved by 
increasing the completeness of the statistical sample and/or reducing
contamination by \OII~emitters relative to the sample 
obtained by the minimum equivalent width requirement. A Bayesian method
takes other observed information into consideration in addition 
to the equivalent width for each $5\sigma$ line detection for the 
purpose of making a classification. This additional information 
comprises the wavelength at which the detected emission line is observed 
($\lambda_{_{\rm EL}}$), the flux measured for the targeted emission line, 
and the detection (or non-detection) of other emission lines expected to 
fall within the spectral range of \HETDEX\ if the emission line is \OII.

\subsubsection{Bayes' Theorem and Classification Threshold}
\label{subsec:bayes}
Bayes' theorem gives the posterior probability of a hypothesis or model $H$ 
given the observed data $B$:
\begin{equation}
  P(H|B) = \frac{P(B|H)P(H)}{P(B)} \,.
\label{eq:bayesthm}
\end{equation}
For a discrete set of $N$ models, the probability of the data $B$ is simply
\begin{equation}
  P(B) = \frac{1}{N} \sum_{j=1}^{N} P(B|H_j) \,.
\label{eq:P_data}
\end{equation}

For two competing models $H_1$ and $H_2$, the {\it posterior odds ratio} is
\begin{equation}
  \frac{P(H_1|B)}{P(H_2|B)} = \frac{P(H_1)}{P(H_2)} \frac{P(B|H_1)}{P(B|H_2)} \,,
\label{eq:posterior_odds_ratio}
\end{equation}
where the first term on the right-hand side is the {\it prior odds ratio} and
the second term, the ratio of the marginal likelihoods under the two models, 
is called the {\it Bayes factor} \citep{Gelman+95,Ivezic+14}. 
The normalization given by Equation~(\ref{eq:P_data}) cancels out.

For our application in the present work, in which we seek to classify galaxies 
detected via a single emission line into two samples, namely, high-redshift LAEs 
(the target sample) and foreground \OII\ emitters (the dominant contaminant), 
the relevant posterior odds ratio is
\begin{equation}
  \frac{P({\rm LAE}|{\rm data})}{P\big({\rm \OII}|{\rm data}\big)} =
  \frac{P({\rm LAE})}{P\big({\rm \OII}\big)}
  \frac{P({\rm data}|{\rm LAE})}{P\big({\rm data}|{\rm \OII}\big)} \,.
\label{eq:applied_posterior_odds_ratio}
\end{equation}
The natural threshold of the posterior odds ratio to classify an emission-line 
detection as LAE is~$>$\,1, but this threshold parameter may be {\it tuned} 
to optimize cosmological results, that is, to minimize variance in $d_A$, 
our estimator of the two-point correlation function.

\subsubsection{Prior Probabilities and Prior Odds Ratio}
\label{subsec:prior_odds}
The prior probability that an object is an LAE or an \OII\ emitter is 
calculated as a function of wavelength, using the luminosity functions 
and cosmological volume elements at the corresponding redshifts. 
For a given model, an observed emission-line wavelength \wlobs\ corresponds 
to exactly one redshift $z$: 
\begin{equation}
\label{eq:redshift}
  \lambda_{_{\rm EL}} = (1+z)\,\lambda \,,
\end{equation}
where $\lambda$ is the line wavelength in the emission rest frame.
The assumed cosmology (given in \S\,\ref{sec:intro}) determines the comoving 
volume $\delta V_c$ of each redshift ``slice'' $\delta z$, corresponding to 
a wavelength interval $\delta \lambda$ in the observed frame.\footnote{Here, 
$\delta \lambda$ denotes a fixed interval of \wlobs\ and not the 
uncertainty of each detected line.}

Under the simplifying assumption that there are only two types of emission-line 
objects, the prior probability that a detected line is \Lya\ and the prior 
probability that it is \OII\ sum up to unity for a given interval $\delta \lambda$:
\begin{equation}
  P({\rm LAE}) + P\big({\rm \OII}\big) = 1 \,.
\label{eq:twotypes}
\end{equation}
Therefore the prior probability that a detected line is \Lya\ is the product of 
the number density at redshift $z$ and the comoving volume at a
corresponding redshift interval $\delta z$:
\begin{equation}
\label{eq:P_LAE}
  P({\rm LAE}) = \frac{n_1 \delta V_1}{n_1 \delta V_1 +n_2 \delta V_2} \,.
\end{equation}
Similarly, the prior probability that a detected line is \OII\ is
\begin{equation}
\label{eq:P_OII}
  P({\rm \OII}) = \frac{n_2 \delta V_2}{n_1 \delta V_1 +n_2 \delta V_2} \,.
\end{equation}

The integral of the Schechter function (Equation~\ref{eq:schechter}) down to the 
line luminosity corresponding to the flux detection limit yields the number densities:
\begin{subequations}
\label{eq:Schechter_int}
\begin{eqnarray}
\label{eq:Schechter_int_LAE}
  n_1 &=&
  \int^{\infty}_{{L^{^{\rm min}}_{_{\rm LAE}}} \big{/} {L^{^{*}}_{_{\rm LAE}}}}
  \phi^{*}_{_{\rm LAE}}
  \left( \tfrac{L}{L^{*}} \right)^{\alpha_{_{\rm LAE}}}
  e^{-L/{L^{*}}} d\left( \tfrac{L}{L^{*}} \right) \,,\\
\label{eq:Schechter_int_OII}
  n_2 &=&
  \int^{\infty}_{{L^{^{\rm min}}_{_{\rm \OII}}} \big{/} {L^{^{*}}_{_{\rm \OII}}}}
  \phi^{*}_{_{\rm \OII}}
  \left( \tfrac{L}{L^{*}} \right)^{\alpha_{_{\rm \OII}}}
  e^{-L/{L^{*}}} d\left( \tfrac{L}{L^{*}} \right) \,,\ \ \ \ 
\end{eqnarray}
\end{subequations}
where the vector of parameters $\bm{\theta} = (\alpha, L^{*}, \phi^{*})$ is described 
in \S\,\ref{subsec:sim-spec} and given in Table~\ref{tab:sim_param}, and the lower limits 
of integration are related to the wavelength-dependent flux detection 
limit $f_{\rm min}$ as follows:
\begin{subequations}
\label{eq:flux_limit}
\begin{eqnarray}
\label{eq:flux_limit_LAE}
  L^{\rm min}_{_{\rm LAE}}(\lambda_{_{\rm EL}}) &=& 4\pi\left[ d_L\big(z_{_{\rm LAE}}(\lambda_{_{\rm EL}})\big) \right]^2 f_{\rm min}(\lambda_{_{\rm EL}}) \,,\\
\label{eq:flux_limit_OII}
  L^{\rm min}_{_{\rm \OII}}(\lambda_{_{\rm EL}}) &=& 4\pi\left[ d_L\big(z_{_{\rm \OII}}(\lambda_{_{\rm EL}})\big) \right]^2 f_{\rm min}(\lambda_{_{\rm EL}}) \,, \ \ 
\end{eqnarray}
\end{subequations}
where $d_L$ is luminosity distance.
The volume of the interval $z \pm \delta z$ is given by the differential 
comoving volume integrated over the redshift interval corresponding to 
$\lambda_{_{\rm EL}} \pm \delta\lambda$:\footnote{Note that $\delta z$, 
which corresponds to a fixed interval $\delta \lambda$ in the observed 
frame (Equation~\ref{eq:redshift}), has different values under the 
assumption of each model.}
\begin{subequations}
\label{eq:vol_int}
\begin{eqnarray}
\label{eq:vol_int_LAE}
  \delta V_1 = \delta V_c(z_{_{\rm LAE}}) &=& \int_{z_{_{\rm LAE}} -\delta z}^{z_{_{\rm LAE}} +\delta z} dV_c \,,\\
\label{eq:vol_int_OII}
  \delta V_2 = \delta V_c(z_{_{\rm \OII}}) &=& \int_{z_{_{\rm \OII}} -\delta z}^{z_{_{\rm \OII}} +\delta z} dV_c \,. \ \ 
\end{eqnarray}
\end{subequations}

The Schechter function integral down to a specified lower limit is 
the upper incomplete gamma function:
\begin{equation}
  \Gamma(s,x) = \int^{\infty}_{x} t^{s-1}\,e^{-t}\,dt \,.
\end{equation}

The prior odds ratio is therefore
\begin{equation}
  \frac{P({\rm LAE})}{P\big({\rm \OII}\big)} = \frac{n_1 \delta V_1}{n_2 \delta V_2} \,.
\label{eq:prior_odds_ratio}
\end{equation}

In the case of \HETDEX, this assumption of Equation~(\ref{eq:twotypes})
is reasonable. Since the VIRUS spectrographs only 
extend to 5500\,\AA, lines such as \Hb\ and the \OIII\ doublet 
will not be observable past $z$\,$\sim$\,0.13 (see 
\S\,\ref{subsec:results-aper-spec}). Moreover, while AGN may produce strong line 
emission at 1549\,\AA\ (\CIV), 1909\,\AA\ (\CIII), and 2798\,\AA\ (\MgII), 
the VIRUS coverage is such that these objects will seldom be single-line 
detections (see \S\,\ref{subsec:disc-other-contam}). By far, the 
dominant source of confusion is between \OII\ and \Lya\
\citep{Adams+11}.

\subsubsection{Likelihood Functions and Bayes Factor}
\label{subsec:likelihood}
The luminosity functions and equivalent width distributions reported by 
\citet{Gronwall+16} for $z$\,=\,2.1 and $z$\,=\,3.1 populations of 
LAEs represent the best current information on the properties 
of LAEs in the \HETDEX\ redshift range. By interpreting these 
distributions as {\it probability density functions}, we can calculate 
the marginal likelihood of measuring the observed data under the assumption 
that the object is an LAE. The marginal likelihood 
of the observation given that the observed object is an \OII\ emitter 
is similarly obtained from the galaxy property distributions for \OII\ 
emitters sampled at 0\,$<$\,$z$\,$<$\,0.56 by \citet{Ciardullo+13}. 
Since any correlation between emission-line luminosity and equivalent width 
is, at best, weak \citep{Cowie+96, Hogg+98, Gronwall+07, Ciardullo+12, Ciardullo+13}, 
we treat these two quantities as statistically independent.

The likelihood function of the observed data $B$ under the assumption of 
model $H_j$ is 
\begin{equation}
  P({B}|{H_j}) = 
  \frac{1}{n_j} \int_{a_{j-}}^{a_{j+}} \Phi_{j}(L)\,dL 
  \int_{b_{j-}}^{b_{j+}} \Psi_{j}(W)\,dW
\end{equation}
where $n_j$ is the normalization of the Schechter function integral,
$j \in \{1,2\}$, as given in Equations~(\ref{eq:Schechter_int}); 
the luminosity function $\Phi$ and the equivalent width 
distribution $\Psi$ are given in 
Equations~(\ref{eq:schechter}) and (\ref{eq:exponential}),
respectively; and the limits of integration are
\begin{eqnarray}
  a_{j\pm} \ &=& \ L_{j}(f_{_{\rm EL}}) \pm \delta L_{j} \,,\\
  b_{j\pm} \ &=& \ W_{j}({\rm EW}_{\rm obs}) \pm \delta W_{j} \,.
\end{eqnarray}
where $\delta L_{j}$ and $\delta W_{j}$ denote small fractional changes 
to the corresponding model quantities, with the fraction of change held 
fixed across the hypotheses.

The exponential form of the equivalent width distribution $\Psi(W)\,dW$ given in 
Equation~(\ref{eq:exponential}) already includes the proper 
normalization.\footnote{The same is true for the lognormal form to be given in 
Equation~(\ref{eq:lognormal}) in \S\,\ref{subsec:disc-imperfect}.}
The luminosity of the emission line is calculated as in 
Equations~(\ref{eq:flux_limit}), but 
with $f_{_{\rm EL}}$ replacing $f_{\rm min}$, i.e.,
\begin{subequations}
\begin{eqnarray}
  L_{_{{\rm Ly}\alpha}}(f_{_{\rm EL}},\lambda_{_{\rm EL}}) = 4\pi\left[ d_L\big(z_{_{\rm LAE}}(\lambda_{_{\rm EL}})\big) \right]^2 f_{_{\rm EL}} \,,\\
  L_{_{\rm \OII}}(f_{_{\rm EL}},\lambda_{_{\rm EL}}) = 4\pi\left[ d_L\big(z_{_{\rm \OII}}(\lambda_{_{\rm EL}})\big) \right]^2 f_{_{\rm EL}} \,.
\end{eqnarray}
\end{subequations}
The Bayes factor used to calculate the posterior odds ratio given 
in Equation~(\ref{eq:applied_posterior_odds_ratio}) is 
\begin{multline}
  \frac{P({\rm data}|{\rm LAE})}{P\big({\rm data}|{\rm \OII}\big)} = \\
  \frac{n_2}{n_1}
  \frac
  {{\int_{L_{_{{\rm Ly}\alpha}}-\delta L}^{L_{_{{\rm Ly}\alpha}}+\delta L} \Phi_{_{\rm LAE}}(L)\,dL}}
  {{\int_{L_{_{\rm \OII}}-\delta L}^{L_{_{\rm \OII}}+\delta L} \Phi_{_{\rm \OII}}(L)\,dL}}
  \frac
  {{\int_{W_{_{{\rm Ly}\alpha}}-\delta W}^{W_{_{{\rm Ly}\alpha}}+\delta W} \Psi_{_{\rm LAE}}(W)\,dW}}
  {{\int_{W_{_{\rm \OII}}-\delta W}^{W_{_{\rm \OII}}+\delta W} \Psi_{_{\rm \OII}}(W)\,dW}} \,.
\label{eq:bayes_factor}
\end{multline}

The present analysis assumes perfect knowledge of luminosity functions and
equivalent width distributions for the simulated populations of
targeted LAEs and low-redshift \OII\ contaminants. While this 
is overly optimistic at present, the initial season of \HETDEX\ data will
measure the luminosity functions and equivalent width distributions of
both populations to high precision. Section \ref{subsec:disc-imperfect} 
explores how this assumption affects our results. In practice, \HETDEX\ 
data will allow for iterative refinement of our Bayesian \textit{priors}, 
as the luminosity functions and equivalent width distributions of LAEs and 
\OII~emitters, and their evolution as functions of redshift, become more 
precisely known.

\section{Results}
\label{sec:results}
\subsection{Cosmological Distance Measurement Uncertainties}
\label{subsec:results-uncertainties}
In order to obtain an indicator of the performance of each classification method, we parametrize 
the fractional uncertainty in angular diameter distance (\dA) measurements as a function of 
contamination and incompleteness of the statistical sample of LAEs, as follows: 
\begin{equation}
\frac{\sigma_{d_{A}}}{d_{A}} =
{A}x_0^{B} +{C}x_1^{D} +{E}x_0^{F}x_1^{G} +{H} \,,
\label{eq:sigma_dA}
\end{equation}
where $x_0 = f_{\rm \OII}$ is fractional contamination (as defined in Equation~\ref{eq:contamfrac})
and $x_1 = 1 - N^{\rm{true\,LAEs}}_{\rm{\,recovered}} \big/ N^{\rm{true\,LAEs}}_{\rm{\,available}}$ 
is sample incompleteness. We use the observable
power spectrum (Equation~\ref{eq:pobscopy}) 
in a \citet{Fisher35} matrix code\footnote{
   \raggedright{\href{http://wwwmpa.mpa-garching.mpg.de/\~komatsu/crl/list-of-routines.html}{\scriptsize\tt
   http://wwwmpa.mpa-garching.mpg.de/\~{}komatsu/crl/list-of- routines.html}}}
\citep{Shoji+09} that marginalizes over the contamination power
spectrum (the second term in Equation~\ref{eq:pobscopy}) to obtain 
$\sigma_{d_{A}}/d_{A}$ for a grid of
contamination and incompleteness values. A linear bias factor of 2.0 
is used for LAEs in both redshift bins.
We then use the grid of results to derive a fitting formula for 
the parameters. The value of 
$\sigma_{d_{A}}/d_{A}$ corresponding to ``perfect'' 
classification in each redshift bin (1.9\,$<$\,$z$\,$<$\,2.5 and
2.5\,$<$\,$z$\,$<$\,3.5) is given by the parameter H 
in Table~\ref{tab:param}.

\begin{table}
\caption{Number of observable LAEs in the main
   \HETDEX\ spring field at nominal flux limits based
   on {\tt Gr16} Schechter functions,
   with simulated spectroscopic measurement noise, and the
   corresponding parameter values for Equation~(\ref{eq:sigma_dA})\\ determined by
   Fisher matrix analysis
\label{tab:param}
}
\vspace{2pt}
\centering
\footnotesize
\begin{spacing}{1.32}
\begin{tabular}{cccc}
\hhline{====}
Redshift bin
 && 1.9\,$<$\,$z$\,$<$\,2.5
 & 2.5\,$<$\,$z$\,$<$\,3.5 \\
LAE counts
 && 446,200 & 396,400 \\
\cline{1-4}

Parameter && \multicolumn{2}{c}{Fisher matrix-derived values} \\
\hhline{-~--}
$A$ && 3.462 & 3.758 \\
$B$ && 1.323 & 0.917 \\
$C$ && 1.359 & 3.539 \\
$D$ && 1.263 & 1.803 \\
$E$ && 11.65 & 8.934 \\
$F$ && 0.775 & 1.099 \\
$G$ && 3.586 & 1.085 \\
$H$ && 1.409 & 1.901 \\
\cline{1-4}
\vspace{1pt}

\end{tabular}
\end{spacing}
\end{table}

Results for Bayesian classification of emission-line galaxies in a
simulated \HETDEX\ catalog are presented for an optimized requirement of
the posterior odds ratio (Equation \ref{eq:applied_posterior_odds_ratio}) 
for selection as LAEs. This requirement minimizes $\sigma_{d_{A}}/d_{A}$ 
given perfect information on the simulated object labels by optimizing 
the trade-off between contamination and incompleteness. Each redshift bin 
may be optimized independently to maximize the total 
amount of information obtained from the full 1.9\,$<$\,$z$\,$<$\,3.5 LAE sample. 
To accomplish this goal, we need to determine a set of values for the eight 
parameters in Equation~(\ref{eq:sigma_dA}) for each redshift bin. In our 
analysis we divided the full spectral range of \HETDEX\ into two redshift 
bins and tuned the required posterior odds ratio for LAE classification 
in each bin separately; Figure~\ref{fig:sample} and Tables~\ref{tab:summary} 
and \ref{tab:compare} present results corresponding to a Bayesian method 
optimized separately for two redshift bins. The value of $\sigma_{d_{A}}/d_{A}$ 
in total is estimated by inverse variance summation.

\subsection{Improvement over Traditional Equivalent Width Method}
\label{subsec:results-improvement}
Compared to the traditional \EWLya\,$>$\,20\,\AA\ narrowband limit to
classify emission-line galaxies as LAEs (which discards all data below
the dashed line in Figure \ref{fig:sample-all}), the Bayesian method 
presented in \S\,\ref{sec:bm} recovers a more complete statistical sample
of high-redshift LAEs without an overall increase in misidentified
low-redshift \OII~emitters. Our Bayesian method is adaptive to prior 
probabilities that reflect the evolution of the galaxy populations and 
the effect of cosmological volume on the relative density of galaxies 
as a function of wavelength.

\begin{figure*}
\centering
\includegraphics[width=\textwidth,trim={0.36in 0.16in 0.0in 0.2in},clip]
{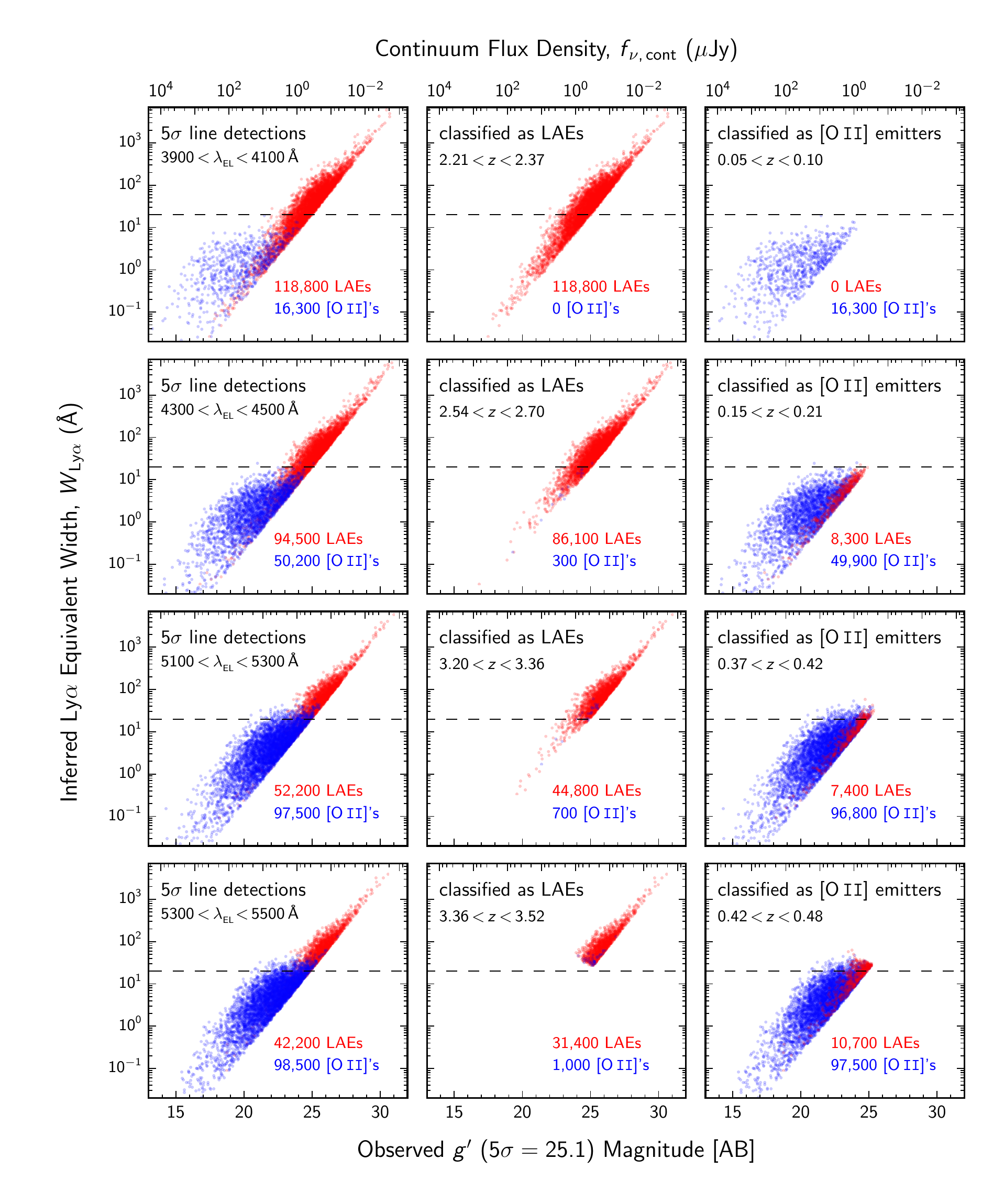}
\caption{
Four selected redshift bins shown in rows for a simulated \HETDEX\ survey with $g'$
(5$\sigma$\,=\,25.1) band imaging survey and realistic
measurement noise. Approximately 5\,\% of the galaxies in each redshift bin are plotted.
\textit{Left}: All $5\sigma$ spectroscopic detections of 
emission-line galaxies whose primary lines are observed in the given 
wavelength interval. \textit{Middle} (\textit{Right}): Simulated 
emission-line galaxies classified by Bayesian method into high-redshift 
LAE (foreground \OII) samples at the corresponding redshifts. Correctly 
classified ``true'' LAEs (\OII~emitters) are shown in red (blue); 
misidentified \OII~emitters (erroneously discarded LAEs) are indicated 
in blue (red).\\
}
\label{fig:sample}
\end{figure*}

LAEs at $z$\,$<$\,2.065 are not contaminated by foreground \OII\
emitters, since \OII\ will not be detected at
$\lambda_{_{\rm{EL}}}$\,$<$\,3727\,\AA. Moreover, at $z$\,$<$\,0.05, 
a galaxy's angular scale is greater than $1\arcsec$ per kpc, hence all 
but the most compact \OII\ sources will be resolved in the imaging survey. 
Consequently, out to about $z$\,$\sim$\,2.4, our Bayesian analysis recovers 
all LAEs with negligible \OII\ contamination 
(see Figure \ref{fig:sample}, top row).

The rate of contamination in the LAE sample identified by
our Bayesian method is sub-percent up to $z$\,$\sim$\,3.0. Over 
1.88\,$<$\,$z$\,$<$\,2.54, the Bayesian method recovers more 
than 99\% of available LAEs, compared 
to a sample identified by the traditional \EWLya\,$>$\,20\,\AA\ 
cutoff that is only $\sim$\,70\% complete.

Table \ref{tab:summary} provides a comparative summary of the two 
classification methods. With respect to the traditional 
\EWLya\,$>$\,20\,\AA\ cut, the Bayesian method significantly increases 
the completeness of the sample of objects classified as LAEs by trading 
near-zero contamination in the case of the EW method for sub-percent 
contamination in the low-redshift bin (1.9\,$<$\,$z$\,$<$\,2.5).

\begin{table*}[t]
\caption{
Classification results for a simulated \HETDEX\ catalog based on {\tt
   Gr16} luminosity functions and \\equivalent width distributions, with
simulated aperture photometry on $g'$ (5$\sigma$\,=\,25.1) band imaging
\label{tab:summary}
}
\vspace{2pt}
\centering
\footnotesize
\begin{spacing}{1.45}
\begin{tabular}{lcccc}
\hhline{=====}

\multirow{2}{*}{Classification method}
   & \multirow{2}{*}{\EWLya\,$>$\,20\,\AA}
   && \multicolumn{2}{c}{{Bayesian method}} \\
\cline{4-5}
   &&& \scriptsize default$^{\rm \,a}$ 
   & \scriptsize {optimized}$^{\rm \,b}$ \\
\cline{1-5}

\ Galaxies classified as LAEs   & 637,400 && 847,500 & {796,200} \\
\cline{1-5}

\ \ \ Missed observable LAEs & 218,700 && 20,100 & {50,800} \\
\ \ \ Sample incompleteness & 26.0\,\% && 2.39\,\% & {6.02\,\%} \\ 
\cline{1-5}

\ \ \ Misidentified \OII~emitters & 13,500 && 25,100 & {4,400} \\
\ \ \ Fractional contamination & 2.12\,\% && 2.96\,\% & {0.55\,\%} \\
\cline{1-5}

\ Measurement uncertainty, $\sigma_{d_{A}}/d_{A}$ & 1.32\,\% && 1.19\,\%
                                           & {1.16\,\%} \\
\cline{1-5}
\vspace{-10pt}\\
\multicolumn{5}{l}{$^{\rm a}$\,\hspace{0.3pt}\scriptsize The 
   ``default'' Bayesian requirement for LAE classification is
   $\tfrac{P(\rm{LAE}|\rm{data})}{P(\rm{\OII}|\rm{data})}$\,$>$\,1.}\\
\multicolumn{5}{l}{$^{\rm b}$\,\hspace{0.3pt}\scriptsize The
   ``optimized'' Bayesian method requires
   $\tfrac{P(\rm{LAE}|\rm{data})}{P(\rm{\OII}|\rm{data})}$\,$>$\,1.38
   for the classification}\\
\multicolumn{5}{l}{\hspace{6pt}\scriptsize of 1.9\,$<$\,$z$\,$<$\,2.5 LAEs and
   $\tfrac{P(\rm{LAE}|\rm{data})}{P(\rm{\OII}|\rm{data})}$\,$>$\,10.3
   for 2.5\,$<$\,$z$\,$<$\,3.5 LAEs.}
\\
\end{tabular}
\end{spacing}
\end{table*}

\begin{table*}
\caption{
Bayesian method classification results for three simulated \HETDEX\
catalogs with $g'$ band imaging ($5\sigma$\,=\,25.1);\\
``baseline'' distributions are used as Bayesian \textit{priors} for 
each simulation scenario\\
\label{tab:compare}
}
\vspace{2pt}
\centering
\footnotesize
\begin{spacing}{1.45}
\begin{tabular}{lccccc}
\hhline{======}

{Simulation scenario} && {baseline} && {pessimistic} & {optimistic} \\

\hhline{------}
\multirow{2}{*}{\ \ \ Distribution of LAEs}
    && \texttt{Gr16} 
    && \texttt{Ci12}, $z$\,=\,2.1 
    & \texttt{Gr07}, $z$\,=\,3.1 \vspace{-2.5pt} \\
             && \scriptsize evolving && \scriptsize no evolution & \scriptsize no evolution \\
\hhline{------}

\ \ \ Distribution of \OII~emitters    && \texttt{Ci13}
    && $+1\sigma \{\phi^{*}, L^{*}, w_{_0}\}_{_{\tt Ci13}}$
    & $-1\sigma \{\phi^{*}, L^{*}, w_{_0}\}_{_{\tt Ci13}}$ \\

\hhline{------}

\ ``True'' observable LAEs    && 842,600 && 375,000 & 1,155,000 \\
\ Galaxies classified as LAEs   && 796,200 && 345,000 & 1,075,800 \\

\hhline{------}

\ \ \ Missed observable LAEs && 50,800 && 32,700 & 83,500 \\
\ \ \ Sample incompleteness && 6.02\,\% && 8.73\,\% & 7.23\,\% \\ 

\hhline{------}

\ \ \ Misidentified \OII~emitters && 4,420 && 7,700 & 4,350 \\
\ \ \ Fractional contamination && 0.55\,\% && 2.20\,\% & 0.40\,\% \\

\hhline{------}
\\

\end{tabular}
\end{spacing}
\end{table*}

Over the entire \HETDEX\ spectral range (3500--5500\,\AA), our 
Bayesian method recovers $\sim$\,25\,\% more LAEs than the
traditional equivalent width method. Over the redshift range
2\,$<$\,$z$\,$<$\,3, 86\,\% of ``true'' LAEs missed by 
\EWLya\,$>$\,20\,\AA\ are correctly classified by the Bayesian method, 
representing the recovery of cosmological information that would be 
discarded by the equivalent width method. In addition to  
improving the completeness of the LAE sample, the Bayesian method 
also reduces contamination by \OII\ sources by a factor of $\sim$\,4 
in our simulations.

\subsection{Aperture Spectroscopy for Additional Emission Lines}
\label{subsec:results-aper-spec}
The presence of other emission lines redward of the primary detected
line (in the case of \OII~emitters; see Figure~\ref{fig:spectra-example}), 
when they fall within the spectral range of \HETDEX, provides additional 
observed information for the two likelihood functions in 
Equation~(\ref{eq:bayes_factor}). Accounting for this spectral 
information leads to better classification performance by the Bayesian 
method in the form of additional reductions in fractional contamination and
further increases in the completeness of the LAE sample.

At $z$\,$<$\,0.1, the vast majority of \OII-emitting galaxies will be
detected via multiple emission lines; we typically observe stronger
\OIII\,$\lambda$5007 emission than \OII\,$\lambda$3727. For the bulk of the
\HETDEX\ \OII\ redshift range, 0.13\,$<$\,$z$\,$<$\,0.42, 
\NeIII\,$\lambda$3869 is the only other line that falls into the 
spectrograph's bandpass.

With our previous assumption of statistically independent quantities, the 
likelihood functions $P\left(\rm{data}|\rm{LAE}\right)$ and 
$P\big(\rm{data}|\rm{\OII}\big)$ are each the product of the 
likelihood functions associated with the individual properties we wish to 
consider:
\begin{equation}
\label{eq:prod_general}
P\left(\rm{data}|\rm{type}\right) 
= \prod_{\rm{properties}}\,P\left(\rm{data}|\rm{type}\right)_{\rm{property}}\,.
\end{equation}
In particular,
\begin{equation}
\label{eq:prod_lines}
P\left(\rm{data}|\rm{type}\right)_{\rm{lines}} 
= \prod_{\rm{lines}}\,P\left(\rm{data}|\rm{type}\right)_{\rm{line}}\,,
\end{equation}
where the lines we wish to consider are \NeIII\,$\lambda$3869, 
\Hb\,$\lambda$4861, \OIII\,$\lambda$4959, and
\OIII\,$\lambda$5007, subject to their falling within the \HETDEX\ 
spectral range. Assuming a Gaussian noise distribution, we can 
calculate the probability of the measured flux at each expected line location. 
When a line is out of range, it contributes no information for or
against the hypothesis that the primary detected line is \OII;
the $P\left(\rm{data}|\rm{type}\right)_{\rm{line}}$ in question is 
set equal to unity, thereby having no effect on the value 
in the left-hand side of Equation~(\ref{eq:prod_lines}).

The improvement due to accounting for additional emission lines is
evident when we consider the boundary at which spectroscopic
information from all additional lines is lost, when
\NeIII\,$\lambda$3869 is redshifted out of the \HETDEX\ spectral
range (3500--5500\,\AA) for $z$\,$>$\,0.42 \OII~emitters
(\wlobs\,$>$\,5299\,\AA). The bottom row in Figure~\ref{fig:sample}
shows $5\sigma$ emission line detections at 5300\,$<$\,\wlobs\,$<$\,5500\,\AA\ 
and their classification into samples of LAEs and \OII~emitters. 
Without spectroscopic information from the additional lines, the Bayesian 
cutoff between LAEs and \OII~emitters is reduced to a nearly straight 
line on a log-log plot of \EWLya\ versus continuum flux densities 
($f_{\nu\rm{,\,cont}}$) in this redshift bin, which is the reddest 200\,\AA\ 
in the spectral range of \HETDEX\ (cf. third row in Figure~\ref{fig:sample}).

\subsection{Optimizing Area versus Depth in Fixed Observing Time}
\label{subsec:results-tradeoff}
\begin{figure*}
\centering
\includegraphics[width=\textwidth,trim={0.36in 0.225in 0.28in 0.2in},clip]
{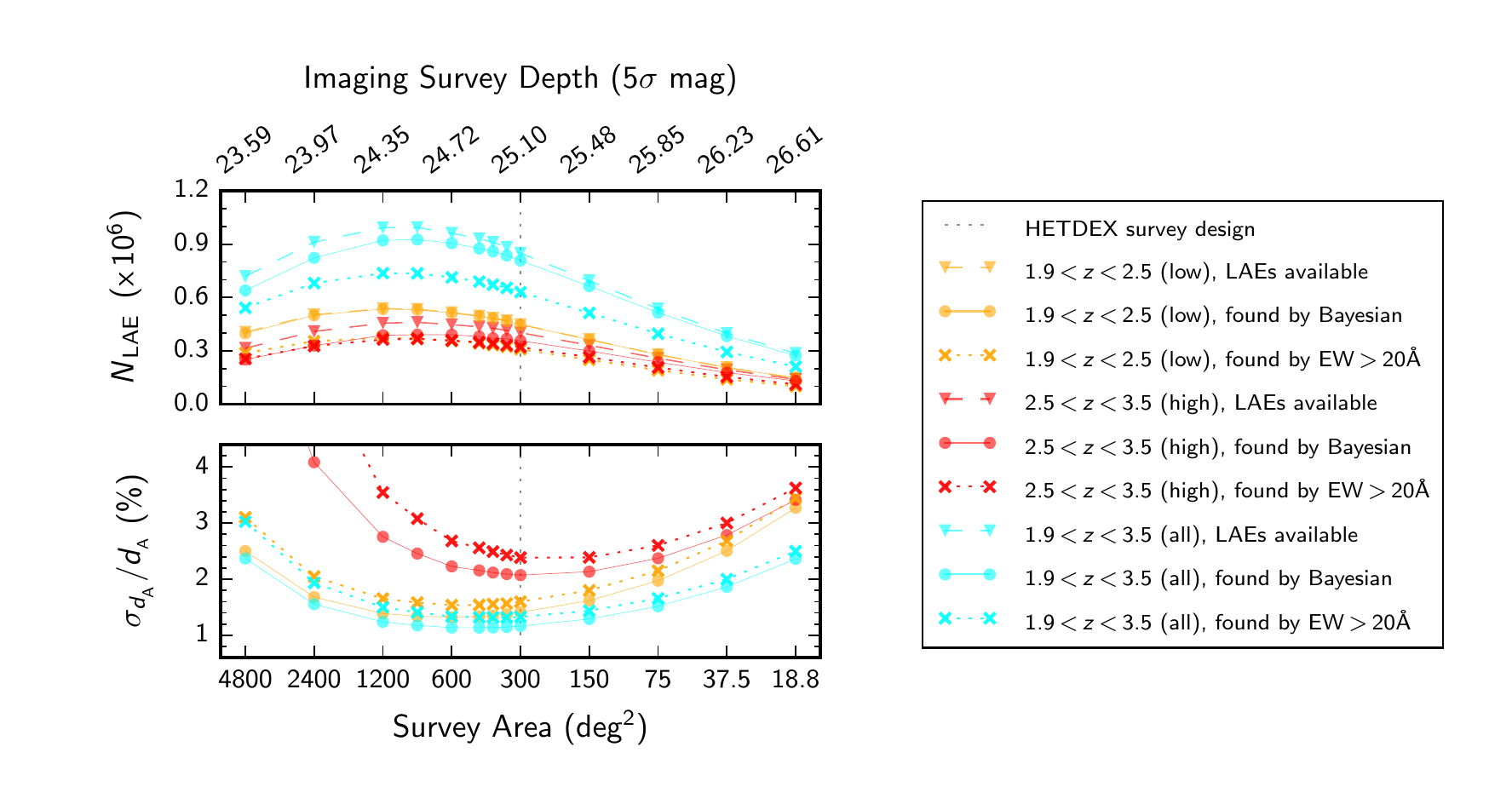}
\caption{
Trade-off between survey area and depth in simulated surveys
for 1.9\,$<$\,$z$\,$<$\,3.5 LAEs in fixed 
broadband imaging and spectroscopic time. A survey that reaches 25.1
magnitude will cover 300~deg$^2$. 
\textit{Top}: Number of LAEs available to be recovered in each
simulated survey and numbers correctly identified by each method for
LAE classification.
\textit{Bottom}: Measurement uncertainty in angular diameter distance (\dA)
corresponding to each combination of classification method and redshift 
range in simulations. The most accurate measurement of \dA\ occurs
with a survey area of 300--600~deg$^2$.\\
}
\label{fig:trade}
\end{figure*}

Using our Bayesian method and \HETDEX\ as a baseline scenario, we
investigate the survey design trade-off between total survey area and 
depth of coverage per unit survey area. Holding the amount of available
observing time fixed at the \HETDEX\ survey design (denoted by the grey
dashed line in Figure~\ref{fig:trade}), we apply $5\sigma$ depths in 
both simulated spectroscopic and imaging surveys that are
modified by $1/\sqrt{t}$, where $t$ is the factor by which observing
time per unit survey area changes as a result of a corresponding change 
in total survey area. Simulated measurement noise varies accordingly, as 
described in \S\,\ref{sec:simulate}.

The number of LAEs available to be recovered in the $5\sigma$ line
flux-limited sample changes with survey design, as shown in the upper
panel of Figure~\ref{fig:trade}. For each survey design, we re-run the 
Fisher matrix code described in \S\,\ref{subsec:results-uncertainties} 
with the number of available LAEs to determine a new set of parameters for 
Equation~(\ref{eq:sigma_dA}) (i.e., Table~\ref{tab:param}) and re-optimize 
the Bayesian method for each case.

Our analysis indicates that the current \HETDEX\ survey design is
effectively optimal in the trade-off between area and depth when our 
Bayesian method is used as the redshift classifier: Trading away from
the nominal 300~deg$^2$ survey area moves the optimal
$\sigma_{d_{A}}/d_{A}$ for the two redshift bins in opposite
directions (lower panel in Figure~\ref{fig:trade}).

\section{Discussion}
\label{sec:disc}

\subsection{Imperfect Knowledge of Distribution Functions of Galaxy Properties}
\label{subsec:disc-imperfect}
The luminosity functions and equivalent width distributions measured 
by \citet{Ciardullo+13} and \citet{Gronwall+16} 
represent the best current information on the galaxy populations 
\HETDEX\ will observe, but will be superseded by data collected in the initial 
season of \HETDEX\ observations. The ability of the Bayesian method to classify 
spectroscopic emission-line detections does not crucially depend on perfect
knowledge of the characteristics of the galaxy populations. To
demonstrate this behavior, we test our method with simulated populations of 
LAEs with varying characteristics, including cases in which the luminosity
function and equivalent width distribution do not evolve with redshift. 
For these tests, the population of foreground \OII~emitters is 
simulated as described in \S\,\ref{sec:simulate}; the Bayesian method 
assumes the \textit{priors} given in \S\,\ref{sec:bm} and uses the
LAE classification cutoff optimized for the ``baseline'' scenario
(Table~\ref{tab:summary}). Table~\ref{tab:compare} compares two such 
test cases with the ``baseline'' scenario.

The $z$\,=\,3.1 \Lya\ luminosity function measured by \citet{Gronwall+07},
if assumed to be constant with redshift and applied to the entire redshift 
range (1.9\,$<$\,$z$\,$<$\,3.5), implies more observable LAEs (whose emission-line 
fluxes exceed the detection limit) at both low and high redshift (see upper-left 
panel of Figure~\ref{fig:sim_param}). This result leads to a lower rate of 
contamination in the LAE sample classified by a Bayesian method that uses 
the same priors as the baseline and which is optimized with respect to the 
baseline scenario. Although the Bayesian method recovers a larger LAE sample 
in this scenario, the sample is more incomplete in fractional terms due to 
the large number of ``true'' observable LAEs available to be recovered.

The $z$\,=\,2.1 \Lya\ luminosity function measured by \citet{Ciardullo+12}, 
if assumed to be constant with redshift and applied to the entire redshift 
range, represents a scenario that is unfavorable for LAE-based cosmological 
study at high redshift. Nevertheless, our Bayesian method misses fewer 
observable LAEs and misidentifies fewer \OII~emitters than does the EW 
method in the baseline scenario.

\citet{Blanton+Lin00} and \citet{Yan+06} demonstrate that the 
\OII\ equivalent width distribution is also well-fit by a lognormal
function. For simulations in which the equivalent widths of \OII\
emitters are lognormally distributed, i.e.,
\begin{equation}
\Psi(W)\,dW = \frac{1}{\sqrt{2\pi}\sigma_{_W}}\,\frac{dW}{W}\,
\exp\left[-\frac{1}{2\sigma_{_W}^2}\left(\ln\frac{W}{W_{_0}}-
\frac{\sigma_{_W}^2}{2}\right)^2\right],
\label{eq:lognormal}
\end{equation}
our Bayesian method recovers similarly robust LAE samples
assuming either exponential or lognormal distributions as priors.
For this test, we fit the \texttt{Ci13} exponential distributions to the form of 
Equation~(\ref{eq:lognormal}) for $W$\,$>$\,5\,\AA. These lognormal 
fits are shown in the lower-right panel of Figure~\ref{fig:sim_param}; 
the fitted parameters are presented in the far-right column of 
Table~\ref{tab:sim_param}.

By accounting for each detected emission line's wavelength, flux, 
equivalent width, and, in the case of most \OII~emitters ($z$\,$<$\,0.42), 
additional lines present in the galaxy spectrum, the Bayesian method's overall 
ability to identify high-redshift LAEs targeted by \HETDEX\ for cosmological study is 
only mildly affected by a mismatch between the expected and actual distributions of galaxy
properties. This reflects a practical situation in which we do not know the 
luminosity functions of the ``real'' populations with high precision at the 
onset of the survey.

\subsection{Sensitivity to Imaging Survey Depth}
\label{subsec:disc-depth}
Since the vast majority of \OII~emitters are brighter in their continua 
than the imaging depth in the survey design of \HETDEX, the Bayesian 
method is able to keep contamination in the sample of objects classified 
as LAEs under 1\,\%, down to survey depth AB\,$\sim$\,22 mag for the 
low-redshift bin~(1.9\,$<$\,$z$\,$<$\,2.5) and AB\,$\sim$\,24 mag for 
the high-redshift bin~(2.5\,$<$\,$z$\,$<$\,3.5).

\begin{figure}
\centering
\includegraphics[width=0.98\linewidth,trim={0.36in 0.25in 0.12in 0.2in},clip]
{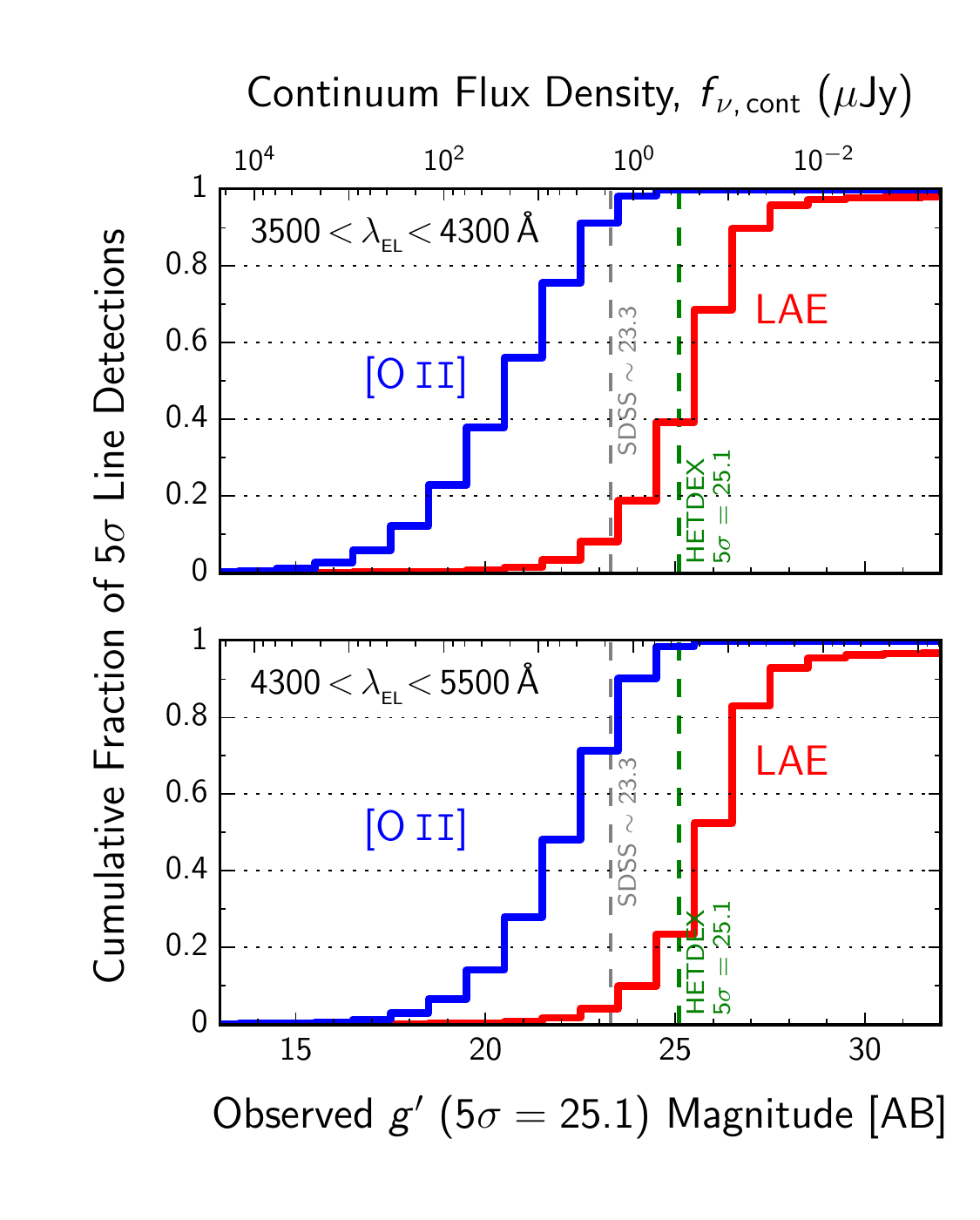}
\caption{
Fractions of $5\sigma$ spectroscopic detections of LAEs and
\OII~emitters, in a simulated \HETDEX\ survey, brighter than the given
observed continuum magnitude of the objects in aperture photometry
on simulated $g'$ band imaging.\\\vspace{2pt}}
\label{fig:cumu}
\end{figure}

Figure \ref{fig:cumu} demonstrates that for the 3500--4300\,\AA\ portion of the
\HETDEX\ spectral range (top panel), essentially all observed \OII\ 
emitters have continuum flux densities greater than the $5\sigma$ limit 
of \HETDEX\ broadband imaging, resulting in small equivalent widths and 
enabling the success of the traditional method for LAE selection in this
regime of observed emission line wavelengths. The inclusion of emission 
line flux in addition to equivalent width in Bayesian classification leads 
to an LAE sample with a rate of contamination less than a quarter percent.

In the longer wavelength bin (4300--5500\,\AA, bottom panel of Figure 
\ref{fig:cumu}), line fluxes measured for \OII~emitters are weaker
than those in the shorter-wavelength bin due to increasing luminosity
distance, rendering \OII\ equivalent widths measured in this bin
similar to those of high-redshift LAEs. 
In this redshift range, accurate classification is relatively difficult for 
both methods. Equivalent width alone is not an adequate classifier; 
the inclusion of line flux and information from additional lines enables 
Bayesian classification to be effective in a regime that poses serious 
challenges to the traditional \EWLya\,$>$\,20\,\AA\ method.

\subsection{Single Broadband Filter with Best Performance}
\label{subsec:disc-bestbroadband}
\begin{figure*}
\centering
\includegraphics[width=0.98\textwidth,trim={0.36in 0.2in 0.0in 0.2in},clip]
{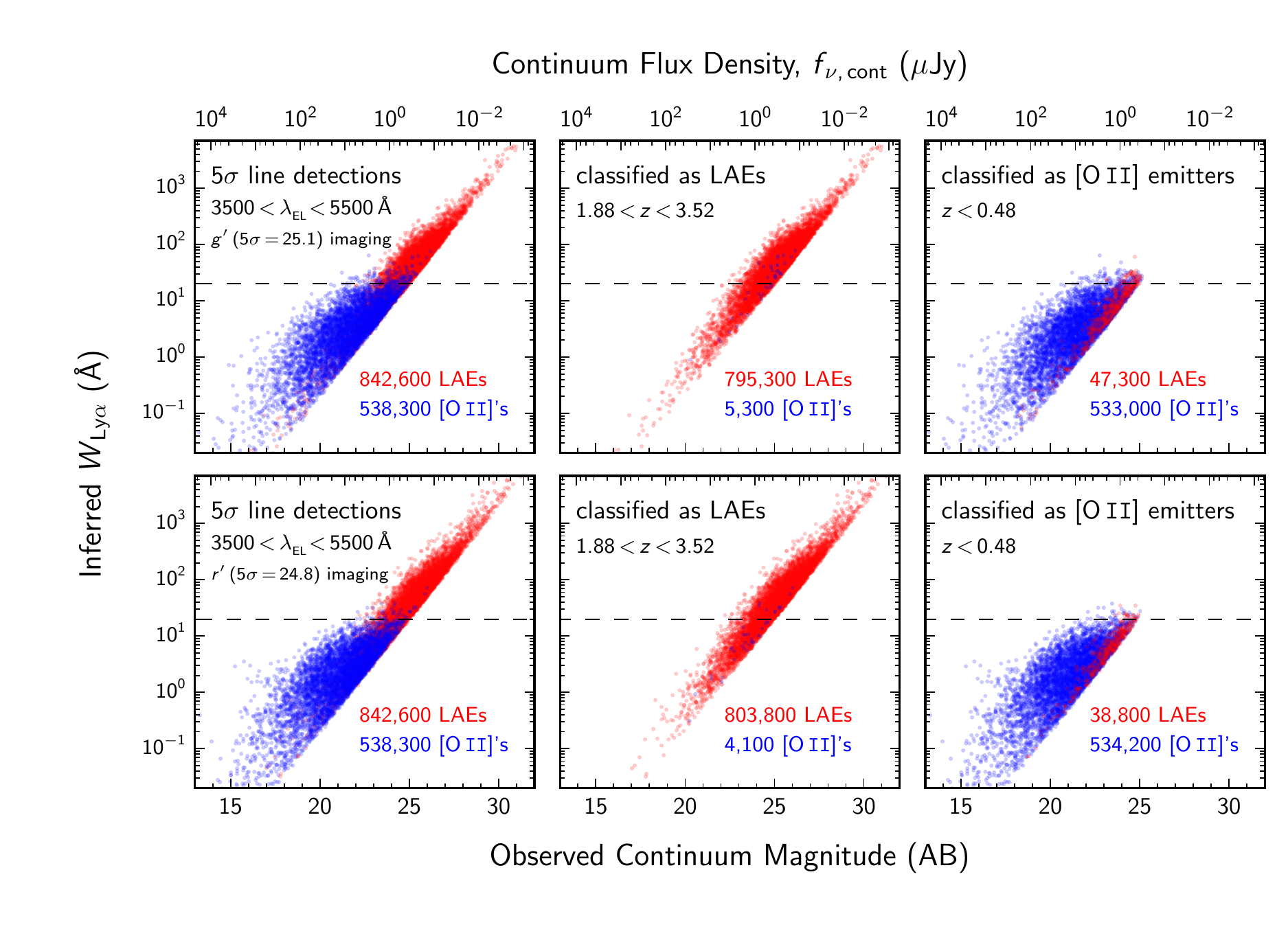}
\caption{
Classification results for a simulated spectroscopic survey coupled with 
$g'$ band imaging (\textit{top}) and $r'$ band imaging
(\textit{bottom}). Approximately 1\,\% of each simulated catalog is
plotted. The distributions of \Lya\ and \OII\ equivalent widths are less similar
in simulated $r'$ band imaging, resulting in a moderate
improvement in classification with respect to the case of $g'$ band
imaging.\\\vspace{12pt}}
\label{fig:compare_sims}
\end{figure*}

The performance of our Bayesian method is similar with $g'$ and $r'$ band 
simulated imaging in equal observing time, assuming a typical 0.3-mag
reduction in equal-time depth going from $g'$ to $r'$. With the Bayesian method, the 
improvement in changing from $g'$ ($5\sigma$\,=\,25.1) to 
$r'$ ($5\sigma$\,=\,24.8) band imaging is a $\sim$\,1\,\% 
reduction in $\sigma_{d_{A}}$. To place this improvement in perspective, 
changing from $g'$ to $r'$ with the equivalent width method reduces $\sigma_{d_{A}}/d_{A}$ 
by $\sim$\,1.5\,\%, i.e., the observed distributions of LAEs and 
\OII~emitters are less similar in $r'$ band imaging. 
Figure~\ref{fig:compare_sims} compares simulated $g'$ and $r'$ band 
imaging for a single realization of a simulated spectroscopic survey.

At higher redshifts (2.5\,$<$\,$z$\,$<$\,3.5), suppressing contamination in 
$g'$ band is problematic for \EWLya\,$>$\,20\,\AA. The relatively 
red colors of \OII~emitters \citep{Bridge+15} lead to weaker continuum flux 
density measurements in the $g'$ band, leading to larger 
photometric equivalent widths for \OII\ determined by aperture photometry 
on $g'$ band imaging (Equation~\ref{eq:EWdef}). This results in a higher rate 
of contamination in the LAE sample selected by the traditional EW method versus 
the case of $r'$ band imaging. In contrast, our Bayesian method is able to 
optimize its requirement for LAE classification given the selection of a 
broadband filter.

\subsection{Splitting Available Observing Time to Obtain Color}
\label{subsec:disc-color}
Inclusion of the colors of galaxies as additional input to the Bayesian
method results in modest further reductions in measurement uncertainty
in angular diameter distance. In 2.5\,$<$\,$z$\,$<$\,3.5, splitting available
observing time between $g'$ and $r'$ in our simulations and using the 
distributions of $g'$$-$$r'$ colors of LAEs and \OII~emitters found by HPS
\citep{Bridge+15} as additional Bayesian \textit{priors} result in 
$\sim$\,5\,\% improvement in $\sigma_{d_{A}}/d_{A}$ over devoting all 
available observing time to obtaining $g'$ band imaging in full (25.1 mag) 
depth. Tests of other broadband filter combinations yield similar results.

\subsection{Other Sources of Contamination}
\label{subsec:disc-other-contam}
Each of the ``additional'' emission lines in \OII\ spectra targeted for 
aperture spectroscopy (\S\,\ref{subsec:results-aper-spec}) may in fact 
be a $5\sigma$ line detection in its own right. However, detections of 
\Hb\,$\lambda$4861, \OIII\,$\lambda$4959, and/or \OIII\,$\lambda$5007 
nearly guarantee that \OII\,$\lambda$3727 will also be detected within 
the \HETDEX\ spectral range. Therefore, \OIII\ will not be a significant 
source of contamination \citep{Adams+11}.

AGNs represent another potential source of confusion for identifying LAEs. 
In most cases, detection of strong \CIV\,$\lambda$1549 provides an 
indication of the source's nature; \CIV\ shifts into the \HETDEX\
spectral range at $z$\,=\,1.25. \CIII\,$\lambda$1909 shifts out of the
red-end of the \HETDEX\ range at $z$\,=\,1.88, but not before
\Lya\,$\lambda$1216 shifts into range at the blue-end of \HETDEX. Hence 
when \CIV\ is observed, we always expect to detect \CIII\ or \Lya.

Spurious emission line detections caused by cosmic rays represent a 
significant source of contamination. Our Bayesian method can
be broadened to account for these spectroscopic detections
that have no counterpart in the imaging survey; doing so will greatly
increase the importance of deep imaging.

\subsection{Application to Future Surveys}
\label{subsec:disc-appl}
\subsubsection{Narrowband Surveys}
In our application (\HETDEX), continuum emission is not well measured in 
spectroscopy for emission line-selected objects, necessitating a 
complementary broadband imaging survey for redshift classification of 
targeted objects. In addition to spectroscopic emission line surveys, 
the Bayesian method presented in this work is equally applicable to 
narrowband surveys.

There are two limiting cases. In the first, where the narrowband filters 
have top-hat transmission profiles \citep[e.g.,][]{Ciardullo+12}, each 
object's line flux will be well-determined, while the precision of its 
redshift measurement is limited to the width of the filter. In the 
second case, the filter transmission curves may be Gaussian 
\citep[e.g.,][]{Gronwall+07}. In this scenario, there is an additional 
probability associated with the location of the emission line within 
the filter bandpass, which creates uncertainty in converting narrowband 
flux density excess into an equivalent width. In either case, ancillary 
broadband data are required to determine the equivalent widths of detected 
emission lines. Redshift classifications can then be made to identify 
galaxies targeted by the survey.

\vspace{12pt}
\subsubsection{{\it Euclid} and WFIRST}
Future surveys by the {\it Euclid} space mission and the space-based 
Wide-Field Infrared Survey Telescope (WFIRST) 
will conduct cosmological studies via BAO measurements with slitless 
spectroscopy for \Ha\ emission-line galaxies at 0.7\,$<$\,$z$\,$<$\,2.1
\citep{Laureijs+11} and 1.3\,$<$\,$z$\,$<$\,2.7
\citep{Green+12}, respectively. Due to their nature in targeting a 
specific line for detection, these types of investigations are generally 
susceptible to contamination by other strong line emissions. For example,
\citet{Geach+08,Geach+10} have found that \OIII\,$\lambda$5007, \Paa, \Pab,
and \FeII\, may all be confused with \Ha\ in low-resolution surveys.

A similar Bayesian method can assist these projects, which will
have many broadbands available. For application to these upcoming
experiments, the method should be broadened to include photometric
redshift probabilities in addition to the luminosity functions and 
equivalent width distributions considered in this study.

When broadband photometric redshifts are available \citep{Pullen+15}, 
they can be combined with our EW-based probabilities to yield a more 
robust classification, via 
$R_{\rm \,total}$\,=\,$R_{\operatorname{photo-{\it z}}}$\,$\times$\,$R_{\rm \,EW}$,
where $R$ refers to posterior odds ratios of LAE versus \OII. However, these
quantities share the information of the continuum flux density near 
the emission-line wavelength, so they are not completely independent, and
hence it would be preferable to perform a joint analysis such as
template-fitting including emission-line information.

\subsection{Use of Classification Probabilities in Large-scale Structure Analyses}
\label{subsec:disc-probs}
The parametrization of $\sigma_{d_{A}}/d_{A}$ given by Equation~(\ref{eq:sigma_dA})
implicitly assumes that $d_{A}$ is produced by way of the power spectrum calculated 
from point estimates of redshift (correspondingly, of object label) based on a cut 
in classification probability. If we instead consider the catalog to comprise observed 
objects each with a known classification probability, we may retain potentially 
valuable information in calculating summary statistics for both populations.

Instead of reducing a probabilistic catalog to a traditional classification problem, 
we may do hierarchical inference directly on the classification probabilities to obtain 
the contamination fraction necessary for calculation of the power spectrum, as outlined in 
Appendix~\ref{apndx:A} \citep[][in prep]{Malz+Hogg16}. An alternative and more ambitious 
approach would convert the classification probabilities to redshift probabilities, thereby 
replacing the density field necessary for the calculation of the power spectrum with one that 
treats each object as having a probability distribution over redshift, a sum of components 
proportional to the classification probabilities at the redshifts corresponding to the  
galaxy types in question. Both of these approaches would preserve the knowledge of the 
classification probabilities but would require greater computational cost, as the 
calculation of the two-point correlation function must take more information into account.

\section{Conclusions}
\label{sec:concl}
The Bayesian method presented in this work for the classification of LAEs
offers robust improvements over the traditional limit requiring LAEs
to have rest-frame equivalent width (\EWLya) greater than 20\,\AA. The
statistical discriminating power of our Bayesian method derives
from cosmological volumes of the corresponding redshifts based on the
assumed cosmology, the properties measured for previously observed samples 
of LAEs and \OII~emitters, and known positions of other emission lines in 
the spectra of \OII~emitters. For a simulated \HETDEX\ catalog with
realistic measurement noise, our Bayesian method:
\begin{itemize}
\vspace{-1.6pt} \item{Recovers 86\,\% of LAEs missed by the
   \EWLya\,$>$\,20\,\AA\ cutoff over 2\,$<$\,$z$\,$<$\,3;}
\vspace{-2.5pt} \item{Outperforms \EWLya\,$>$\,20\,\AA\ in limiting 
   contamination in the LAE sample and increases the completeness 
   of the statistical sample;}
\vspace{-2.5pt} \item{Allows trade-off between contamination and 
   incompleteness in arbitrary wavelength/redshift bins.}
\end{itemize}

For simulated \HETDEX\ catalogs, Table~\ref{tab:summary} shows that our 
implementation of the Bayesian method reduces
uncertainties in angular diameter distance measurements by 14\,\%, which
is equivalent to obtaining 29\,\% more data, compared to the
\EWLya\,$>$\,20\,\AA\ criterion.\\

Additional conclusions of our investigation are:
\begin{itemize}
\vspace{-1.6pt} \item{For fixed spectroscopic depths, performance of
   the Bayesian method is relatively insensitive to imaging survey
   depth, suggesting that maximizing imaging survey 
   area should be favored in a fixed amount of observing time for the purpose 
   of LAE-\OII\ galaxy separation;}
\vspace{-2.5pt} \item{Inclusion of the colors of galaxies as an input to the 
   Bayesian method increases discriminating power and results in modest 
   further reductions in distance errors;}
\vspace{-2.5pt} \item{The Bayesian method can also be used to determine 
   which single broadband filter produces the best performance;}
\vspace{-2.5pt} \item{The Bayesian method can be directly applied to other 
   surveys where single emission lines require classification, including 
   planned space-based observations by {\it Euclid} and WFIRST.}
\end{itemize}

Unlike the Bayesian approach, machine learning methods do not require
prior assumptions on the luminosity functions and equivalent width
distributions of galaxies, but they require a sizable training set,
consisting of ancillary data for which the object labels 
are known for 5 to 10\,\% of the survey \citep{Acquaviva+14}. As a 
result, we anticipate the combination of these two complementary 
classification approaches to yield additional improvements.

\acknowledgments
We thank the referee for providing critical comments and 
detailed suggestions, which have led to substantial improvement in the clarity of 
the method presented in this paper. E.\,G. and A.\,S.\,L. gratefully acknowledge 
support from the NSF through Grant AST-1055919. E.\,K. is supported in part by 
MEXT KAKENHI Grant Number 15H05896. J.\,J.\,F. thanks the University of Texas at 
Austin for hospitality and support during a sabbatical visit, where a portion of 
this work was completed.
\vspace{6pt}

\bibliographystyle{apj}
\bibliography{myrefs}
\vspace{2pt}

\appendix
\section{A. \ The Effect of Contamination by \OII~Emitters}
\label{apndx:A}
When \OII\,$\lambda$3727 lines are misidentified as \Lya\,$\lambda$1216 lines,
we erroneously map the correlation function of \OII\ galaxies at low
redshifts to the correlation function of LAEs at high redshifts. As a
result, instead of measuring 
the correlation at some comoving separations between LAEs at $z$\,$>$\,1.9, we
actually measure the correlation of \OII~emitters at {smaller}
comoving separations at $z$\,$<$\,0.5.

This effect can be formulated most straightforwardly using real space correlation
functions. Let us imagine that, from the \HETDEX\ data, we measured the
correlation function, $\xi(\bm{\theta}d_A,s_{\parallel})$, where
$\bm{\theta}$ is the angular separation on the sky, $d_A$ is the
{comoving} angular diameter distance to a given redshift (assigned to LAEs, 
including misidentified \OII~emitters), and $s_{\parallel}$ is the line-of-sight
separation. 

A problem arises because since we misidentified the \OII\ lines for
\Lya\ lines, our assumed values of $d_A$ and $s_\parallel$ are
incorrect. 

Let us denote the correlation function of the contamination at 
$z$\,$>$\,1.9 as $\xi_{\rm con}$, and the true correlation function of the 
\OII~emitters at $z$\,$<$\,0.5 as $\xi_{\rm \OII}$. As these are the same
objects, except for the values of the spatial coordinates assigned to each, 
we have a trivial relation: 
\begin{equation}
 \xi_{\rm con}\left(\bm{\theta}d_A,s_{\parallel}\right) \,
= \, \xi_{\rm \OII}\left(\bm{\theta}d_A^{\rm \OII},s_{\parallel}^{\rm \OII}\right).
\end{equation}
The angular separation, $\bm{\theta}$, is the same
for both $\xi_{\rm con}$ and $\xi_{\rm \OII}$. The misidentification of
the lines simply produces incorrect values for $d_A$ and $s_{\parallel}$ on the
left-hand side.

The power spectrum can be obtained from the correlation function 
via the usual inverse Fourier transform:
\begin{equation}
 P_{\rm con}\left(\bm{k}_\perp,k_{\parallel}\right)
= 
\int d^2\bm{s}_\perp \, e^{-i\bm{k}_\perp \bm{s}_{\perp}}
\int ds_{\parallel} \, e^{-ik_\parallel s_\parallel} \,
\xi_{\rm con}\left(\bm{s}_\perp,s_{\parallel}\right),
\label{eq:pobs}
\end{equation}
where $\bm{s}_\perp\equiv \bm{\theta}d_A$.
We also Fourier transform $\xi_{\rm con}\left(\bm{s}_\perp,s_{\parallel}\right)$:
\begin{equation}
 \xi_{\rm con}\left(\bm{s}_\perp,s_{\parallel}\right)
= \xi_{\rm \OII}\left(\bm{s}^{\rm \OII}_\perp,s_{\parallel}^{\rm \OII}\right)
=
\int\frac{d^2\bm{k}^{\rm \OII}_\perp}{(2\pi)^2} \, 
e^{i\bm{k}^{\rm \OII}_\perp \bm{s}^{\rm \OII}_{\perp}}
\int\frac{dk^{\rm \OII}_{\parallel}}{2\pi} \, 
e^{ik^{\rm \OII}_\parallel s^{\rm \OII}_\parallel} \, P_{\rm \,\OII}\left(k^{\rm \OII}\right).
\label{eq:xicon}
\end{equation}

Now, define the key parameters, the ratios of $s_\perp$ and
$s_\parallel$, as follows:
\begin{eqnarray}
\label{eq:alpha}
 \alpha \, &\equiv& \, \frac{s_\perp}{s_\perp^{\rm \OII}}
= \frac{d_A}{d_A^{\rm \OII}} > 1,\\
\label{eq:beta}
\beta \, &\equiv& \, \frac{s_\parallel}{s_\parallel^{\rm \OII}}
= \frac{(1+z)/H(z)}{(1+z_{\rm \,\OII})/H(z_{\rm \,\OII})}
= \frac{3727}{1216}\frac{H(z_{\rm \,\OII})}{H(z)}
= \frac{3727}{1216}\sqrt{\frac{\Omega_m(1+z_{\rm
    \,\OII})^3+\Omega_\Lambda}{\Omega_m(1+z)^3+\Omega_\Lambda}}.
\end{eqnarray}
The quantity $\alpha$ is always larger than 1 (i.e., the redshift of LAEs
is always higher than that of \OII~emitters), while $\beta$ can be
larger or smaller than 1. Of course, $1+z_{\rm \,\OII}$ satisfies the relation:
\begin{equation}
 1+z_{\rm \,\OII} = (1+z)\,\frac{1216}{3727} \simeq \frac{1+z}{3.065}.
\end{equation} 
In Table~\ref{tab:alphabeta}, we summarize the values of $\alpha$ and
$\beta$ for representative redshifts: $z=$ 2.2 and 3.0.

\begin{table}[h]
\caption{
Values of $\alpha$ and $\beta$
\label{tab:alphabeta}
}
\vspace{2pt}
\centering
\footnotesize
\begin{spacing}{1.45}
\begin{tabular}{cccccc}
\hhline{======}
{$z$}
& {$z_{\rm \,\OII}$}
& {$d_A(z)$$^{\rm \,a}$}
& {$d_A(z_{\rm \,\OII})$$^{\rm \,a}$}
& {$\alpha$$^{\rm \,b}$}
& {$\beta$$^{\rm \,c}$} \\

\hhline{------}
2.2 & 0.044 & 3,900 & 130.7 & 29.8 & 1.008 \\
3.0 & 0.305 & 4,554 & 853.3 & 5.34 & 0.833 \\
\hhline{------}
\multicolumn{4}{l}{$^{\rm a}$\,In units of $h^{-1}$\,Mpc} && \\
\multicolumn{4}{l}{$^{\rm b}$\,$\alpha\equiv \frac{d_A(z)}{d_A\left(z_{\rm
   \,\OII}\right)}$} && \vspace{2pt}\\
\multicolumn{4}{l}{$^{\rm c}$\,$\beta\equiv \frac{(1+z)H\left(z_{\rm \,\OII}\right)}{\left(1+z_{\rm
 \,\OII}\right)H(z)}$} && \\

\end{tabular}
\end{spacing}
\end{table}

Using Equation~(\ref{eq:xicon}) in Equation~(\ref{eq:pobs}) along with
Equations~(\ref{eq:alpha}) and (\ref{eq:beta}), we find
\begin{eqnarray}
 \nonumber
P_{\rm \,con}\left(\bm{k}_\perp,k_{\parallel}\right)
&=&
\int\frac{d^2\bm{k}^{\rm \OII}_\perp}{(2\pi)^2}
\int d^2\bm{s}_\perp \, e^{i\Big(\frac{\bm{k}^{\rm
			   \OII}_\perp}{\alpha}-\bm{k}_\perp\Big)\bm{s}_{\perp}}
\int\frac{dk^{\rm \OII}_{\parallel}}{2\pi}
\int ds_{\parallel} \, e^{i\Big(\frac{k^{\rm
			 \OII}_\parallel}{\beta}-k_\parallel\Big)\bm{s}_\parallel}
\times P_{\rm \,\OII}\left(k^{\rm \OII}\right)\\
&=&
(\alpha^2\beta)\,P_{\rm \,\OII}\left(\sqrt{\alpha^2k_\perp^2+\beta^2k_\parallel^2}\right).
\label{eq:pkcon}
\end{eqnarray}
This is the equation we use for computing the contamination of the LAE
power spectrum. This result makes physical sense:
\begin{itemize}
 \item[1.] For a given set of values of $k_\perp$ and $k_\parallel$, we
	    are actually observing the power spectrum of \OII~emitters
	    at smaller scales, $k_\perp\to \alpha k_\perp$ and
	    $k_\parallel\to \beta k_\parallel$. This contamination produces a {horizontal}
	    shift of the \OII\ power spectrum to smaller $k$ values.
 \item[2.] As the correlation function measures the dimensionless power
	    spectrum, $k_\perp^2 k_\parallel P(k)$, the normalization of
	    the power spectrum is also shifted by $\alpha^2\beta$,
             generating a {vertical} shift of the \OII\ power spectrum.
\end{itemize}

At lower redshifts, $\alpha$ can be as large as $30$ (see Table~\ref{tab:alphabeta}),
which boosts the amplitude of the \OII\ power spectrum by a factor of
$\alpha^2=900$. Conversely, $\beta$ is of order unity for the
redshift range of interest.

Finally, the observed power spectrum is given by the weighted average of
the LAE power spectrum, $P_{\rm \,LAE}$, and the contamination power
spectrum:
\begin{equation}
 P_{\rm{\,obs}}\left(k_\perp,k_\parallel\right)
= {\left(1-f_{\rm \OII}\right)^2}\,
P_{\rm{\,LAE}}\left(\sqrt{k_\perp^2+k_\parallel^2}\right)
   + {f_{\rm \OII}^2} ({\alpha^2\beta})\,
P_{\rm{\,\OII}}\left(\sqrt{\alpha^2k_\perp^2+\beta^2k_\parallel^2}\right),
\label{eq:pobsfinal}
\end{equation}
where $f_{\rm \,\OII}$ is the fraction of \OII~emitters in the total
sample, i.e., 
\begin{equation}
 f_{\rm \,\OII}\equiv \frac{\mbox{number of contaminating \OII\
   emitters}}{\mbox{number of galaxies classified as LAEs}}.
\end{equation}
The contamination (the second term in Equation~\ref{eq:pobsfinal})
makes the observed power spectrum {\it anisotropic}, even in the
absence of redshift space distortions and the Alcock-Paczy\'nski 
effect \citep{Komatsu10}.

\end{document}